\documentclass[aps,prd,twocolumn,preprintnumbers,floatfix,reprint,nofootinbib]{revtex4}
\usepackage{graphicx}
\usepackage{dcolumn}
\usepackage{amsmath}
\usepackage{bm}
\usepackage{hyperref}
\usepackage{subfigure}
\usepackage{multirow}
\usepackage{subfigure}
\usepackage{rotating}
\usepackage{color}
\usepackage{multirow}
\usepackage{dcolumn}

\usepackage{longtable,lscape}
\usepackage{overpic}
\usepackage{booktabs}
\usepackage{makecell}
\usepackage{diagbox}

\newcolumntype{C}[1]{>{\centering\arraybackslash}p{#1}}

\renewcommand\arraystretch{1.8}
\allowdisplaybreaks

\begin{document}
\title{ Exotic pentaquark states with the $qqQQ\bar{Q}$ configuration}
\author{Hong-Tao An$^{1,2}$}
\email{anht14@lzu.edu.cn}
\author{Qin-Song Zhou$^{1,2}$}
\author{Zhan-Wei Liu$^{1,2}$}
\email{liuzhanwei@lzu.edu.cn}
\author{Yan-Rui Liu$^3$}
\email{yrliu@sdu.edu.cn}
\author{Xiang Liu$^{1,2}$}
\email{xiangliu@lzu.edu.cn}
\affiliation{$^1$
School of Physical Science and Technology, Lanzhou University, Lanzhou 730000, China\\
$^2$Research Center for Hadron and CSR Physics, Lanzhou University
and Institute of Modern Physics of CAS, Lanzhou 730000, China\\
$^3$School of Physics,
Shandong University, Jinan 250100, China}

\date{\today}
\begin{abstract}
In the framework of the color-magnetic interaction model, we have systematically calculated the mass splittings for the S-wave triply heavy pentaquark states with the configuration $qqQQ\bar{Q}$ $(Q=c,b;q=u,d,s)$. Their masses are estimated and their stabilities are discussed according to possible rearrangement decay patterns. Our results indicate that there may exist several stable or narrow such states. We hope the present study can help experimentalists to search for exotic pentaquarks.
\end{abstract}
\maketitle

\section{Introduction}\label{sec1}

The possible existence of mutiquark states beyond the ordinary hadrons were first proposed by M. Gell-Mann and G. Zweig \cite{GellMann:1964nj,Zweig:1981pd}. Nowadays, it is still an important and interesting topic to look for such states \cite{Liu:2019zoy}. With the experimental progress in recent decades, we are able to find heavy quark multiquark candidates in various processes. In fact, experimentalists announced more exotic states in past years since the Belle Collaboration reported the observation of X(3872) in 2003 \cite{Choi:2003ue,Choi:2007wga,Mizuk:2008me,Mizuk:2009da,Liu:2013dau,Xiao:2013iha,Ablikim:2013mio,Ablikim:2013wzq,Ablikim:2013xfr,Chilikin:2014bkk,Aaij:2014jqa,Ablikim:2015gda,Ablikim:2015swa,Ablikim:2015tbp},
which provides good opportunities to study the nonperturbative color interactions. Some of the states have been considered as good tetraquark candidates \cite{Swanson:2006st,Zhu:2007wz,Voloshin:2007dx,Drenska:2010kg,Chen:2016qju,Hosaka:2016pey,Richard:2016eis,Lebed:2016hpi,Esposito:2016noz}.

Recently, the LHCb Collaboration reported the observation of new pentaquark states at the Rencontres de Moriond QCD conference \cite{LHCbtalk,Aaij:2019vzc}. By analyzing the $J/\psi p$ invariant mass spectrum in the $\Lambda_b^0$ decay with updated data, a new pentaquark $P_c(4312)$ was discovered with $7.3\sigma$ significance. Meanwhile, the analysis shows two narrow subpeaks with 5.4 significance, $P_c(4440)$ and $P_c(4457)$, for the previously reported $P_c(4450)$ \cite{Aaij:2015tga,Aaij:2016ymb}.
Although within the framework of one-boson-exchange (OBE) model these three new states could be identified clearly as loosely bound $\Sigma_c\bar{D}$ molecule with $I(J^P)=1/2(1/2^-)$, $\Sigma_c\bar{D}^*$ with $I(J^P)=1/2(1/2^-)$, and $\Sigma_c\bar{D}^*$ with $I(J^P)=1/2(3/2^-)$, respectively \cite{Chen:2019asm,Chen:2019bip,Liu:2019tjn,He:2019ify,Huang:2019jlf,Xiao:2019mvs,Shimizu:2019ptd,Guo:2019kdc}, Ram\'{\i}rez \textit{at al.} in Ref. \cite{Fernandez-Ramirez:2019koa} find the evidence that the attractive interaction in the $\Sigma_{c}^{+}\bar{D}$ channel is not strong enough for forming a bound state.

Because of complicated interactions between the internal quarks, generally, it is hard to distinguish whether a hadron is a tightly bound tetraquark (pentaquark) state, a conventional meson (baryon), a molecular state, or a structure in others pictures. In understanding the internal structures of $P_{c}(4380)$ and $P_{c}(4450)$, many interpretations were proposed, such as the $\Sigma_{c}\bar{D}^{*}$, $\Sigma_{c}^{*}\bar{D}$, and $\Sigma_{c}^{*}\bar{D}^{*}$ molecules \cite{Chen:2015loa,Roca:2015dva,He:2015cea,Ortega:2016syt,Yamaguchi:2016ote,He:2016pfa,Burns:2015dwa,Huang:2015uda,Chen:2015moa,Chen:2016otp,Shimizu:2016rrd,Shen:2016tzq},
compact pentaquark states \cite{Eides:2015dtr,Perevalova:2016dln,Santopinto:2016pkp,Deng:2016rus,Takeuchi:2016ejt},
diquark-diquark-antiquark states, diquark-triquark states \cite{Maiani:2015vwa,Anisovich:2015cia,Li:2015gta,Ghosh:2015ksa,Wang:2015ava,Wang:2015epa,Lebed:2015tna,Zhu:2015bba},
$\bar{D}$ solitons, and kinematical effects from the triangle singularity or due to $\chi_{c1}p$ rescattering \cite{Mikhasenko:2015vca,Liu:2015fea,Guo:2016bkl,Scoccola:2015nia,Guo:2015umn}.

The studies of more possible pentaquarks were also stimulated by the observation of $P_{c}$ states \cite{Lebed:2015dca,Yang:2015bmv,Anisovich:2015zqa,Wang:2016dzu,Chen:2016ryt}. After the experimental confirmation of the doubly charmed baryon $\Xi_{cc}^{++}$ \cite{Aaij:2017ueg,Mattson:2002vu}, the multiquark states with two or more heavy quarks were studied in many works \cite{Lebed:2015dca,Yang:2015bmv,Anisovich:2015zqa,Wang:2016dzu,Chen:2016ryt,Chen:2018cqz}. For example, two possible triple-charm molecular pentaquarks $\Xi_{cc}D_{1}$ and $\Xi_{cc}D_{2}^{*}$ were considered in Ref. \cite{Wang:2019aoc}. In this paper, we systematically study the mass splittings of compact pentaquark states with the $qqQQ\bar{Q}$ configuration ($q=u, d, s; Q=c,b$). If a heavy quark-antiquark pair forms an unflavored state, such pentaquarks look like excited $qqQ$ baryons. Otherwise, they are explicitly exotic states. At present, it is still not easy to dynamically solve the multi-body problem. Here, we use the color-magnetic interaction (CMI) model to calculate the mass splittings and investigate the mass spectrum of the $qqQQ\bar{Q}$ pantaquark states preliminarily. One may consult relevant studies with other methods in Refs. \cite{Hofmann:2005sw,Chen:2017jjn,Wang:2018ihk}.

The Hamiltonian of the quark potential model consists of the one-gluon-exchange interaction part and non-perturbative scalar confining part, which was proposed by de Rujula, Georgi, and Glashow in Ref. \cite{DeRujula:1975qlm}. For the ground state hadrons with the same quark content, such as $\Delta$ and $N$, their mass splitting is mainly determined by the color-magnetic interaction \cite{Karliner:2014gca}. When the spacial contributions are encoded into effective quark masses and coupling parameters, the Hamiltonian can be written as the form containing just the quark mass term and the color-spin interaction term and one gets the CMI model. There are many studies about the mass spectrum for  multiquark systems within this model \cite{Wu:2016vtq, Wu:2016gas, Chen:2016ont, Wu:2017weo, Luo:2017eub, Zhou:2018pcv, Li:2018vhp, Wu:2018xdi,Park:2018oib,Weng:2019ynv,Richard:2019fms}.
The qualitative properties of the obtained spectra are helpful for us to search for relevant exotic states. In the early stage studies on the pentaquark properties, color-magnetic effects were intensively considered as the primary contribution in an attempt to explain the narrow hadronic resonances, too \cite{Montanet:1980te}.

This paper is organized as follows. In Sec. \ref{sec2}, we introduce the CMI model and construct the $flavor\otimes color \otimes spin$ wave functions for the $qqQQ\bar{Q}$ pentaquark states. In Sec. \ref{sec3}, we calculate the relevant Hamiltonian elements and present the corresponding results. In Sec. \ref{sec4}, we give numerical results for the masses of the pentaquark states, illustrate their possible rearrangement decay channels, and discuss the stability of the states. Finally, we present a summary in Sec. \ref{sec5} and an appendix in Sec. \ref{sec6}.

\section{The color-magnetic interaction and the wave functions}\label{sec2}

The Hamiltonian of the CMI model has a simple form
\begin{eqnarray}\label{Eq1}
     H&=&\sum_i^5M_i+H_{\rm CMI},\nonumber\\
     H_{\rm CMI}&=&-\sum_{i<j}C_{ij} \vec\lambda_i\cdot \vec\lambda_j\vec\sigma_i\cdot\vec\sigma_j.\nonumber
\end{eqnarray}
Here, $M_i$ represents the effective quark mass for the $i$-th quark or antiquark and it takes account of effects from kinetic energy, color confinement, and other terms in realistic potential models. The effective constant $C_{ij}\sim\langle\delta^{3}(\vec{r}_{ij})\rangle/(m_im_j)$ reflects the coupling strength between the $i$-th quark and the $j$-th quark, which depends on the quark masses and the spatial wave functions of the ground states \cite{Zhou:2018pcv}.
The Pauli matrix $\sigma_i$ and Gell-Mann matrix $\lambda_i(-\lambda_i^{*})$ act on the spin and color wave functions of the $i$-th quark (antiquark), respectively.

To calculate the required matrix elements, we construct the wave functions of the ground $qqQQ\bar{Q}$ pentaquark states. They are the direct products of $\rm SU(3)_f$ flavor wave function,  $\rm SU(3)_c$ color wave function, and $\rm SU(2)_s$ spin wave function. Here, we treat the heavy quark/antiquark as a flavor singlet state instead of constructing the wave function with flavor $\rm SU(4)_f$ symmetry \cite{Yuan:2012wz}. It is convenient to adopt the diquark-diquark-antiquark base in organizing the wave functions. The notion ``diquark'' only means two quarks and the meaning is different from that in the diquark model in Ref. \cite{Maiani:2004vq} where the diquark is a strongly correlated quark-quark substructure with color=${\bar3}_c$ and spin=$0$. The constructed wave functions may also be used to study properties of the $qqQQ\bar{Q}$ states in dynamical quark models.

\begin{figure}[h]\centering
\includegraphics[width=200pt]{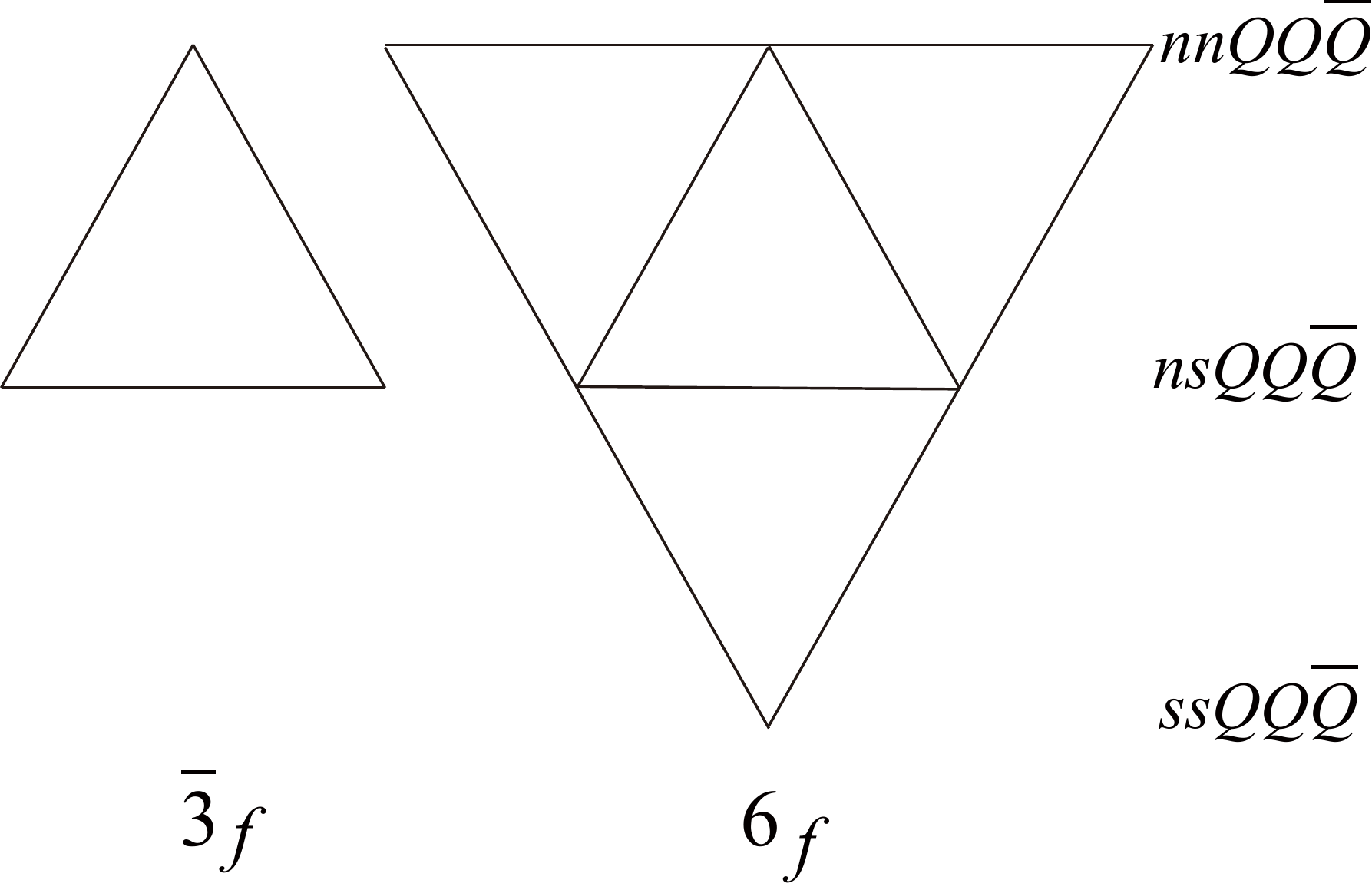}\\
\caption{Flavor representations of the $qqQQ\bar{Q}$ pentaquark states.
Here $Q=c, b$ and $q=n, s$ with $n=u, d$.}\label{fig1}
\end{figure}

In the $\rm SU(3)_f$ flavor space, the $qqQQ\bar{Q}$ states belong to the flavor symmetric $6_f$ and antisymmetric $\bar3_f$ representations (Fig. \ref{fig1}), which is similar to the situation for part of the $QQqq\bar{q}$ states \cite{Zhou:2018pcv}. For the $nnQQ\bar{Q}$ ($n=u,d$) case, the isovector states ($I=1$) and the isoscalar states ($I=0$) do not mix since we do not consider isospin breaking effects. For the $nsQQ\bar{Q}$ case, the fact $m_{n}\neq m_{s}$ leads to $\rm SU(3)_f$ breaking and thus the state mixing between $6_f$ and $\bar3_f$. As a result, we need to consider four cases of states: $nnQQ\bar{Q}$ ($I=1$), $nnQQ\bar{Q}$ ($I=0$), $nsQQ\bar{Q}$ ($I=1/2$), and $ssQQ\bar{Q}$ ($I=0$). Note that the isovector and isoscalar $nnQQ\bar{Q}$ states are not degenerate since the Pauli principle has impacts.

In the color space, the wave functions can be analyzed with the $\rm SU(3)_c$ group theory \cite{Kaeding:1995vq}. The Young diagrams tell us that there are three color-singlet wave functions for the $qqQQ\bar{Q}$ states. With the diquark-diquark-antiquark base, they are
\begin{eqnarray}\label{Eq2}
\phi_1\equiv\phi^{AA}&=&[(q_{1}q_{2})^{\bar{3}_c}(Q_{3}Q_{4})^{\bar{3}_c}\bar{Q}],\nonumber\\
\phi_2\equiv\phi^{AS}&=&[(q_{1}q_{2})^{\bar{3}_c}(Q_{3}Q_{4})^{6_c}\bar{Q}],\nonumber \\
\phi_3\equiv\phi^{SA}&=&[(q_{1}q_{2})^{6_c}(Q_{3}Q_{4})^{\bar{3}_c}\bar{Q}].\nonumber
\end{eqnarray}
In the notation $[(q_{1}q_{2})^{color1}(Q_{3}Q_{4})^{color2}\bar{Q}]$, the $color1$ and $color2$ stand for the color representations of the light diquark and heavy diquark, respectively. The $S$ ($A$) means ``symmetric'' (``antisymmetric'') with quark exchanges. The explicit wave functions are the same as those for the $QQqq\bar{q}$ states studied in Ref. \cite{Zhou:2018pcv}.


One can also use the baryon-meson base ($qqQ$-$Q\bar{Q}$ or $qQQ$-$q\bar{Q}$) to construct the wave functions. The relevant decomposition is
\begin{eqnarray}\label{Eq2}
&&(3_{c}\otimes3_{c}\otimes3_{c})\otimes(3_{c}\otimes\bar{3}_{c})\nonumber\\
&=&(1_{c}\oplus8_{c}\oplus8_{c}\oplus10_{c})\otimes(1_{c}\oplus8_{c})\nonumber\\
&\to&(1_{c}\otimes1_{c})\oplus(8_{c}\otimes8_{c})\oplus(8_{c}\otimes8_{c}).
\end{eqnarray}
Ref. \cite{Wu:2017weo} adopted this base in studying the hidden-charm pentaquark states. Although the final Hamiltonians are different for these two bases, the eigenvalues and mass spectrum would be identical after diagonalization. However, the baryon-meson base is not suitable to the present systems since two pairs of identical quarks may exist in a state like $nncc\bar{Q}$.

\begin{table*}[ht]
\caption{The possible color-spin wave function bases.}\label{bases}
\begin{tabular}{|lll|}
\bottomrule[1.5pt]
\bottomrule[0.5pt]
\multicolumn{3}{|l|}{$J^p={\frac52}^{-}$:}\\
$\phi_{1}\chi_{1}=[(q_{1}q_{2})^{\bar{3}}_1(Q_{3}Q_{4})^{\bar{3}}_1\bar{Q}]^\frac52_2 D^A_{12}$&$\quad$
$\phi_{2}\chi_{1}=[(q_{1}q_{2})^{\bar{3}}_1(Q_{3}Q_{4})^{6}_1\bar{Q}]^\frac52_2 D^A_{12}D_{34}$ &$\quad$
$\phi_{3}\chi_{1}=[(q_{1}q_{2})^{6}_1(Q_{3}Q_{4})^{\bar{3}}_1\bar{Q}]^\frac52_2 D^S_{12}$\\
\bottomrule[0.7pt]
\multicolumn{3}{|l|}{$J^p={\frac32}^{-}$:}\\
$\phi_{1}\chi_{2}=[(q_{1}q_{2})^{\bar{3}}_1(Q_{3}Q_{4})^{\bar{3}}_1\bar{Q}]^\frac32_2 D^A_{12}$&$\quad$
$\phi_{2}\chi_{2}=[(q_{1}q_{2})^{\bar{3}}_1(Q_{3}Q_{4})^{6}_1\bar{Q}]^\frac32_2 D^A_{12}D_{34}$&$\quad$
$\phi_{3}\chi_{2}=[(q_{1}q_{2})^{6}_1(Q_{3}Q_{4})^{\bar{3}}_1\bar{Q}]^\frac32_2 D^S_{12}$
\\
$\phi_{1}\chi_{3}=[(q_{1}q_{2})^{\bar{3}}_1(Q_{3}Q_{4})^{\bar{3}}_1\bar{Q}]^\frac32_1 D^A_{12}$&$\quad$
$\phi_{2}\chi_{3}=[(q_{1}q_{2})^{\bar{3}}_1(Q_{3}Q_{4})^{6}_1\bar{Q}]^\frac32_1 D^A_{12}D_{34}$&$\quad$
$\phi_{3}\chi_{3}=[(q_{1}q_{2})^{6}_1(Q_{3}Q_{4})^{\bar{3}}_1\bar{Q}]^\frac32_1 D^S_{12}$
\\
$\phi_{1}\chi_{4}=[(q_{1}q_{2})^{\bar{3}}_1(Q_{3}Q_{4})^{\bar{3}}_0\bar{Q}]^\frac32_1 D^A_{12}D_{34}$&$\quad$
$\phi_{2}\chi_{4}=[(q_{1}q_{2})^{\bar{3}}_1(Q_{3}Q_{4})^{6}_0\bar{Q}]^\frac32_1 D^A_{12}$&$\quad$
$\phi_{3}\chi_{4}=[(q_{1}q_{2})^{6}_1(Q_{3}Q_{4})^{\bar{3}}_0\bar{Q}]^\frac32_1 D^S_{12}D_{34}$
\\
$\phi_{1}\chi_{5}=[(q_{1}q_{2})^{\bar{3}}_0(Q_{3}Q_{4})^{\bar{3}}_1\bar{Q}]^\frac32_1 D^S_{12}$&$\quad$
$\phi_{2}\chi_{5}=[(q_{1}q_{2})^{\bar{3}}_0(Q_{3}Q_{4})^{6}_1\bar{Q}]^\frac32_1 D^S_{12}D_{34}$&$\quad$
$\phi_{3}\chi_{5}=[(q_{1}q_{2})^{6}_0(Q_{3}Q_{4})^{\bar{3}}_1\bar{Q}]^\frac32_1 D^A_{12}$
\\
\bottomrule[0.7pt]
\multicolumn{3}{|l|}{$J^p={\frac12}^{-}$:}\\
$\phi_{1}\chi_{6}=[(q_{1}q_{2})^{\bar{3}}_1(Q_{3}Q_{4})^{\bar{3}}_1\bar{Q}]^\frac12_1 D^A_{12}$&$\quad$
$\phi_{2}\chi_{6}=[(q_{1}q_{2})^{\bar{3}}_1(Q_{3}Q_{4})^{6}_1\bar{Q}]^\frac12_1 D^A_{12}D_{34}$&$\quad$
$\phi_{3}\chi_{6}=[(q_{1}q_{2})^{6}_1(Q_{3}Q_{4})^{\bar{3}}_1\bar{Q}]^\frac12_1 D^S_{12}$
\\
$\phi_{1}\chi_{7}=[(q_{1}q_{2})^{\bar{3}}_1(Q_{3}Q_{4})^{\bar{3}}_1\bar{Q}]^\frac12_0 D^A_{12}$&$\quad$
$\phi_{2}\chi_{7}=[(q_{1}q_{2})^{\bar{3}}_1(Q_{3}Q_{4})^{6}_1\bar{Q}]^\frac12_0 D^A_{12}D_{34}$&$\quad$
$\phi_{3}\chi_{7}=[(q_{1}q_{2})^{6}_1(Q_{3}Q_{4})^{\bar{3}}_1\bar{Q}]^\frac12_0 D^S_{12}$
\\
$\phi_{1}\chi_{8}=[(q_{1}q_{2})^{\bar{3}}_1(Q_{3}Q_{4})^{\bar{3}}_0\bar{Q}]^\frac12_1 D^A_{12}D_{34}$&$\quad$
$\phi_{2}\chi_{8}=[(q_{1}q_{2})^{\bar{3}}_1(Q_{3}Q_{4})^{6}_0\bar{Q}]^\frac12_1 D^A_{12}$&$\quad$
$\phi_{3}\chi_{8}=[(q_{1}q_{2})^{6}_1(Q_{3}Q_{4})^{\bar{3}}_0\bar{Q}]^\frac12_1 D^S_{12} D_{34}$
\\
$\phi_{1}\chi_{9}=[(q_{1}q_{2})^{\bar{3}}_0(Q_{3}Q_{4})^{\bar{3}}_1\bar{Q}]^\frac12_1 D^S_{12}$&$\quad$
$\phi_{2}\chi_{9}=[(q_{1}q_{2})^{\bar{3}}_0(Q_{3}Q_{4})^{6}_1\bar{Q}]^\frac12_1 D^S_{12}D_{34}$&$\quad$
$\phi_{3}\chi_{9}=[(q_{1}q_{2})^{6}_0(Q_{3}Q_{4})^{\bar{3}}_1\bar{Q}]^\frac12_1 D^A_{12}$
\\
$\phi_{1}\chi_{10}=[(q_{1}q_{2})^{\bar{3}}_0(Q_{3}Q_{4})^{\bar{3}}_0\bar{Q}]^\frac12_0 D^S_{12}D_{34}$&$\quad$
$\phi_{2}\chi_{10}=[(q_{1}q_{2})^{\bar{3}}_0(Q_{3}Q_{4})^{6}_0\bar{Q}]^\frac12_0 D^S_{12}$&$\quad$
$\phi_{3}\chi_{10}=[(q_{1}q_{2})^{6}_0(Q_{3}Q_{4})^{\bar{3}}_0\bar{Q}]^\frac12_0 D^A_{12}D_{34}$
\\
\bottomrule[0.5pt]
\bottomrule[1.5pt]
\end{tabular}
\end{table*}

In the spin space, the possible wave functions for the considered states in the diquark-diquark-antiquark base are
\begin{eqnarray}
J^P=\frac{5}{2}^-:&&\chi_{1}=[(q_{1}q_{2})_{1}(Q_{3}Q_{4})_{1}\bar{Q}]_{2}^{\frac52}, \nonumber \\
J^P=\frac32^-:&&\left\{\begin{array}{ccc}
\chi_{2}&=&[(q_{1}q_{2})_{1}(Q_{3}Q_{4})_{1}\bar{Q}]_{2}^{\frac{3}{2}},\\
\chi_{3}&=&[(q_{1}q_{2})_{1}(Q_{3}Q_{4})_{1}\bar{Q}]_{1}^{\frac{3}{2}},\\
\chi_{4}&=&[(q_{1}q_{2})_{1}(Q_{3}Q_{4})_{0}\bar{Q}]_{1}^{\frac{3}{2}},\\
\chi_{5}&=&[(q_{1}q_{2})_{0}(Q_{3}Q_{4})_{1}\bar{Q}]_{1}^{\frac{3}{2}}, \nonumber \\
\end{array}\right.
\\
J^P=\frac12^-:&&\left\{\begin{array}{ccc}
\chi_{6}&=&[(q_{1}q_{2})_{1}(Q_{3}Q_{4})_{1}\bar{Q}]_{1}^{\frac{1}{2}},\\
\chi_{7}&=&[(q_{1}q_{2})_{1}(Q_{3}Q_{4})_{1}\bar{Q}]_{0}^{\frac{1}{2}},\\
\chi_{8}&=&[(q_{1}q_{2})_{1}(Q_{3}Q_{4})_{0}\bar{Q}]_{1}^{\frac{1}{2}},\\
\chi_{9}&=&[(q_{1}q_{2})_{0}(Q_{3}Q_{4})_{1}\bar{Q}]_{1}^{\frac{1}{2}},\\
\chi_{10}&=&[(q_{1}q_{2})_{0}(Q_{3}Q_{4})_{0}\bar{Q}]_{0}^{\frac{1}{2}}.
\end{array}\right.\nonumber
\end{eqnarray}
In the notation $[(q_{1}q_{2})_{spin1}(Q_{3}Q_{4})_{spin2}\bar{Q}]^{spin4}_{spin3}$, $spin_1$ and $spin_2$ represent the spins of the light and heavy diquarks, respectively, $spin_3$ represents the total spin of the four quarks, and $spin_4$ represents the total spin of the pentaquark. The diquark is symmetric (antisymmetric) when $spin_{1,2}$ is $1$ ($0$).


Combining the spin and color wave functions together, we obtain thirty possible bases which are shown in Table \ref{bases} with the notation $[(q_{1}q_{2})^{color1}_{spin1}(Q_{3}Q_{4})^{color2}_{spin2}\bar{Q}]^{spin4}_{spin3}$. Not all of them are allowed for a given set of quantum numbers.
To reflect the constraint from the Pauli principle, we have inserted three symbols $D^A_{12}$, $D^S_{12}$, and $D_{34}$ in the wave functions. When the light diquark is symmetric (or antisymmetric) in flavor space, $D^S_{12}=0$ (or $D^A_{12}=0$), otherwise $D^S_{12}=1$ (or $D^A_{12}=1)$. When the two heavy quarks are the same, $D_{34}=0$, otherwise $D_{34}=1$. Considering all possible configurations, we need to analyze twelve $qqQQ\bar{Q}$ systems. They can be divided into six classes:\\
(1) $nncc\bar{Q}$ $(I=1)$,
$nnbb\bar{Q}$ $(I=1)$,
$sscc\bar{Q}$,
$ssbb\bar{Q}$;\\
(2) $nncc\bar{Q}$ $(I=0)$,
$nnbb\bar{Q}$ $(I=0)$;\\
(3) $nncb\bar{Q}$ $(I=1)$,
$sscb\bar{Q}$;\\
(4) $nncb\bar{Q}$ $(I=0)$;\\
(5) $nscc\bar{Q}$,
$nsbb\bar{Q}$;\\
(6) $nscb\bar{Q}$.\\
Each class has similar structures and the same CMI Hamiltonian expressions.

\section{The CMI Hamiltonian expressions}\label{sec3}
With the constructed wave functions, we can calculate CMI Hamiltonian matrix elements.
To simplify the expressions, we define the combinations of the effective couplings shown in Table \ref{variable substitution}.

\begin{table}[htbp]
\caption{Defined variables to simplify the CMI Hamiltonian matrix elements.}\label{variable substitution} \centering
\begin{tabular}{cc|cc}
\bottomrule[1.5pt]
\bottomrule[0.5pt]
Variable & Definition&Variable & Definition\\
\midrule[1pt]
$\alpha$&$C_{12}+C_{34}$&$\zeta$&$C_{13}+C_{23}+C_{14}+C_{24}$\\
$\beta$&$C_{34}-2C_{12}$&$\eta$&$C_{13}+C_{23}-C_{14}-C_{24}$\\
$\gamma$&$2C_{34}-C_{12}$ &$\theta$&$C_{13}-C_{23}+C_{14}-C_{24}$ \\
$\delta$&$3C_{34}-C_{12}$&$\kappa$&$C_{13}-C_{23}-C_{14}+C_{24}$ \\
$\epsilon$&$C_{34}-3C_{12}$&&  \\
$\nu$&$C_{15}+C_{25}$&$\xi$&$C_{15}-C_{25}$\\
$\lambda$&$C_{35}+C_{45}$&$\mu$&$C_{35}-C_{45}$\\
\bottomrule[0.5pt]
\midrule[1.5pt]
\end{tabular}
\end{table}

For the pentaquark states without constraints from the Pauli principle, e.g. $nscb\bar{Q}$, all the color-spin wave function bases in Table \ref{bases} are involved. In the Appendix, we show the obtained CMI matrices for the cases $J^P=5/2^-$, $3/2^-$, and $1/2^-$ in Tables \ref{11}, \ref{12}, and \ref{13}, respectively. For the pentaquark states having constraints from the Pauli Principle, relevant matrices can be extracted from these tables. Here, we take the $nncc\bar{Q}$ case as an example. 
When one considers the $I(J^{P})=1({5/2}^-)$ state, one has $D^{S}_{12}=0$, $D^{A}_{12}=1$, and $D_{34}=0$ and only the base $\phi_{1}\chi_{1}$ is allowed. It is easy to read out the CMI Hamiltonian from Table \ref{11},
\begin{eqnarray}\scriptsize
\langle H_{\rm CMI}\rangle= \frac{1}{3}
[8\alpha+2\zeta+4(\nu+\lambda)].
\end{eqnarray}
Similarly, when one considers the $I(J^{P})=0({5/2}^-)$ state, only the wave function base $\phi_{3}\chi_{1}$ is allowed because $D^A_{12}=0$, $D^S_{12}=1$, and $D_{34}=0$. The extracted CMI Hamiltonian from Table \ref{11} is
\begin{eqnarray}\label{eq14}
\langle H_{\rm CMI}\rangle=\frac{1}{3}(4\gamma+5\zeta-2\lambda+10\nu).
\end{eqnarray}

\section{The $qqQQ\bar{Q}$ pentaquark mass spectra}\label{sec4}
\subsection{The determination of parameters and estimation strategy}

Now, we determine the values of the seventeen coupling parameters ($C_{nn}$, $C_{ns}$, $C_{ss}$, $C_{cn}$, $C_{bn}$, $C_{cs}$, $C_{bs}$, $C_{bc}$, $C_{cc}$, $C_{bb}$, $C_{n\bar{c}}$, $C_{n\bar{b}}$, $C_{s\bar{c}}$, $C_{s\bar{b}}$, $C_{c\bar{c}}$, $C_{b\bar{c}}$, $C_{b\bar{b}}$, and $C_{c\bar{b}}$) in order to estimate the pentaquark masses.
Most of them can be extracted from the measured masses of the conventional hadrons (see Table \ref{comp}). The related CMI expressions are
\begin{eqnarray}\label{12}
H_{\rm CMI}^{J=1}(q_1\bar{q}_2)&=&\frac{16}{3}C_{12},\nonumber\\
H_{\rm CMI}^{J=0}(q_1\bar{q}_2)&=&-16C_{12},\nonumber\\
H_{\rm CMI}^{J=\frac32}(q_1q_2q_3)&=&\frac{8}{3}(C_{12}+C_{23}+C_{13}),\nonumber\\
H_{\rm CMI}^{J=\frac12}(q_1q_2q_3)&=&\frac{8}{3}\begin{pmatrix}
C_{12}-2[C_{23}+C_{13}]&\sqrt3[C_{23}-C_{13}]\nonumber \\
\sqrt3[C_{23}-C_{13}]&-3C_{12}\nonumber
\end{pmatrix},
\end{eqnarray}
where the two bases for the last matrix corresponds to the case of $J_{q_1q_2}=1$ and that of $J_{q_1q_2}=0$.
The obtained coupling parameters have been listed in Table I of Ref. \cite{Liu:2019zoy} and we just use these coupling parameters to calculate CMI Hamiltonian.

\begin{table}[htbp]
\caption{Used masses of the conventional hadrons in units of MeV \cite{Tanabashi:2018oca}. The adopted masses of the not-yet-observed doubly heavy baryons are taken from Ref. \cite{Gershtein:2000nx}. The values in parentheses are obtained with the parameters in Ref. \cite{Li:2018vhp}.}\label{comp}
\begin{tabular}{cc|cc|cc|cc}
\bottomrule[1.5pt]
\bottomrule[0.5pt]
\multicolumn{2}{c|}{Mesons}&\multicolumn{2}{|c|}{Mesons}&\multicolumn{2}{|c|}{Baryons}&\multicolumn{2}{|c}{Baryons}\\ \multicolumn{2}{c|}{($J=0$)}&\multicolumn{2}{|c|}{($J=1$)}&\multicolumn{2}{|c|}{($J=\frac12$)}&\multicolumn{2}{|c}{($J=\frac32$)}\\
\bottomrule[0.5pt]
$D$&1869.7&$D^{*}$&2010.3&$\Sigma_c$&2454.0&$\Sigma_c^*$&2518.4\\
$D_s$&1968.3&$D_{s}^{*}$&2112.2&$\Xi_c^{'}$&2577.9&$\Xi_c^*$&2645.5\\
&&&&$\Omega_{c}$&2695.2&$\Omega_{c}^{*}$&2765.9\\
$B$&5279.3&$B^{*}$&5324.7&$\Sigma_{b}$&5811.3&$\Sigma_{b}^*$&5832.1\\
$B_s$&5366.9&$B_s^*$&5415.4& $\Xi_b^{'}$&5935.0&$\Xi_b^*$&5955.3\\
&&&&$\Omega_{b}$&6046.4&$\Omega_{b}^{*}$&6090.0\cite{Yin:2019bxe}\\
$\eta_{c}$&2983.9&$J/\psi$&3096.9& $\Xi_{cc}$&3621.4&$\Xi_{cc}^*$&(3685.4)\\
&&&&$\Xi_{cc}$&(3557.4)&$\Xi_{cc}^*$&3621.4\\
&&&&$\Omega_{cc}$&(3730.4)&$\Omega_{cc}^*$&3802.4\cite{Weng:2018mmf}\\
$\eta_{b}$&9399.0&$\Upsilon$&9460.3&$\Xi_{bb}$&10093.0&$\Xi_{bb}^*$&(10113.8)\\
&&&&$\Omega_{bb}$&10193.0&$\Omega^{*}_{bb}$&(10212.2)\\
$B_c$&6275.1&$B^{*}_{c}$&6331.0\cite{Mathur:2018epb}&$\Xi_{bc}$&6820.0&&\\
&&&&$\Xi^{'}_{bc}$&(6845.9)&$\Xi_{bc}^*$&(6878.8)\\
&&&&$\Omega_{bc}$&6920.0&&\\
&&&&$\Omega^{'}_{bc}$&(6950.9)&$\Omega_{bc}^*$&(6983.4)\\
\bottomrule[0.5pt]
\midrule[1.5pt]
\end{tabular}
\end{table}

\begin{table}[htbp]
\caption{Mass differences ($\Delta M=M_{Th.}-M_{Ex.}$) between the calculated values (Th.) and experimental values (Ex.) for conventional hadrons in units of MeV. $M_{Th.}$ is obtained with $M=\sum_i M_i +\langle H_{\rm CMI}\rangle$ and $M_n=361.7$ MeV, $M_s=540.3$ MeV, $M_c=1724.6$ MeV, and $M_b=5052.8$ MeV \cite{Li:2018vhp}.}\label{theoreticalvalue}
\begin{tabular}{cc|cc|cc|cc}
\bottomrule[1.5pt]
\bottomrule[0.5pt]
Hadron&$\Delta M$&Hadron&$\Delta M$&Hadron&$\Delta M$&Hadron&$\Delta M$\\
\bottomrule[0.5pt]
$\pi$&109.5&$\rho$&107.2&  $N$&0&$\Delta$&0\\
$K$  &110.6&$K^*$&105.3&   $\Sigma$&-12.4&$\Sigma^*$&-5.4\\

$\omega$&99.8&$\phi$&96.0 &       $\Xi$&9.4&$\Xi^*$&-7.3\\
$D$&112.2&$D^{*}$&113.7&   $\Lambda$&1.1&$\Omega$&0\\
$D_s$&189.7&$D_{s}^{*}$&188.7&  $\Sigma_{c}$&0&$\Sigma_{c}^*$&0\\
$B$&101.6&$B^{*}$&101.2 &       $\Lambda_{c}$&14.7& $\Xi_{c}$&58.4\\
$B_s$&189.6&$B_s^*$&190.2&      $\Xi'_{c}$&35&$\Xi_{c}^*$&37.6\\
$\eta_{c}$&380.9&$J/\psi$&381.0&  $\Omega_c$&76.5&$\Omega_c^*$&82.6\\
$\eta_{b}$&660.0&$\Upsilon$&661.0&  $\Sigma_b$&0&$\Sigma_{b}^{*}$&0\\
$B_c$&450.0 &&&        $\Lambda_b$&9.7& $\Xi_b$&62.7\\
&&&&$\Xi'_{b}$&39.8&$\Xi_b^*$&45.0\\
&&&&$\Omega_b$&92.1 & $\Xi_{cc}$&161.5\\
\bottomrule[0.5pt]
\midrule[1.5pt]
\end{tabular}
\end{table}

Using the mass formula $M=\sum_i M_i +\langle H_{\rm CMI}\rangle$ and the obtained parameters, one sees that the estimated masses of conventional hadrons are in general higher than the measured values, which is illustrated in Table \ref{theoreticalvalue}. The reason is that the adopted model and parameters could not account for the necessary attractions for all the hadrons. Overestimated masses with this approach were also obtained in various tetraquark and pentaquark states \cite{Wu:2016vtq,Chen:2016ont,Luo:2017eub,Wu:2017weo,Li:2018vhp,Zhou:2018pcv,Wu:2016gas,Wu:2018xdi}. To make a more reasonable estimation, we use the improved mass formula by replacing $\sum_i M_i$ in Eq. (\ref{Eq1}) with $M_{ref}-\langle H_{\rm CMI}\rangle_{ref}$ where $M_{ref}$ is a reference mass scale and $\langle H_{\rm CMI}\rangle$ is the corresponding CMI matrix element. Then
\begin{eqnarray}\label{Eq11}
M=M_{ref}-\langle H_{\rm CMI}\rangle_{ref}+\langle H_{\rm CMI}\rangle.
\end{eqnarray}
In the present study for pentaquark states, we choose the baryon-meson thresholds as the mass scales, where the reference baryon-meson system should have the same constituent quarks with a considered system. The attraction not incorporated in the original approach is somehow phenomenologically compensated in this procedure \cite{Zhou:2018pcv}.

Before the detailed discussions about the $qqQQ\bar{Q}$ pentaquark states, we emphasize that our results are only rough estimations. They should be updated once a $qqQQ\bar{Q}$ pentaquark state is observed in future experiments and its mass can be chosen as a reference scale. Although the pentaquark masses may be changed largely, the mass splittings should not be affected significantly.

In the following parts, we only present the numerical values obtained with Eq. (\ref{Eq11}). Here, the involved masses of reference baryons and mesons have been given in Table \ref{comp}. To understand the decay properties in the following discussions, we also show some masses of the not-yet-observed doubly heavy baryons in the table, which were obtained from several theoretical calculations. Since the spin of the $\Xi_{cc}$ observed by LHCb may be 1/2 or 3/2, we show results in both cases in Table \ref{comp}.

\subsection{The $nncc\bar{Q}$, $sscc\bar{Q}$, $nnbb\bar{Q}$, and $ssbb\bar{Q}$ pentaquark states}

\begin{table}[htbp]
\centering \caption{The estimated masses for the $nncc\bar{Q}$ $(Q = c, b)$ $(I=1, 0)$ system in units of MeV.
The values in the second column in each case are eigenvalues of the CMI Hamiltonian and those after this column are determined with the relevant reference systems.
}\label{mass-ssccQ}
\resizebox{0.96\columnwidth}{!}{
 \begin{tabular}{cccc|ccc}
\bottomrule[1.5pt]
\bottomrule[0.5pt]
\multicolumn{4}{l|}{$nncc\bar {c}(I=1)$}&\multicolumn{3}{l}{$nncc\bar {c}(I=0)$}\\
$J^p$&Eigenvalue&$(\Sigma_{c}J/\psi)$ &$(\Xi_{cc}\bar{D})$&Eigenvalue&$(\Sigma_{c}J/\psi)$ &$(\Xi_{cc}\bar{D})$\\
\bottomrule[0.7pt]
$\frac52^-$&100.5&5616.7&5732.7&
48.5&5564.7&5680.7\\
$\frac{3}{2}^{-}$ &
$\begin{pmatrix}117.3\\81.2\\56.9\\-43.0\end{pmatrix}$&
$\begin{pmatrix}5633.5\\5597.4\\5573.2\\5473.2\end{pmatrix}$&
$\begin{pmatrix}5744.1\\5708.0\\5683.8\\5583.8\end{pmatrix}$&
$\begin{pmatrix}36.5\\-68.8\\-161.0\end{pmatrix}$&
$\begin{pmatrix}5552.7\\5447.4\\5355.2\end{pmatrix}$&
$\begin{pmatrix}5668.7\\5563.4\\5471.2\end{pmatrix}$\\
$\frac{1}{2}^{-}$ &
$\begin{pmatrix}182.8\\104.4\\3.9\\-62.7\end{pmatrix}$&
$\begin{pmatrix}5699.1\\5620.6\\5520.1\\5453.5\end{pmatrix}$&
$\begin{pmatrix}5815.0\\5736.5\\5636.0\\5569.5\end{pmatrix}$&
$\begin{pmatrix}12.8\\-71.0\\-124.3\\-267.2\end{pmatrix}$&
$\begin{pmatrix}5529.0\\5445.2\\5391.9\\5249.0\end{pmatrix}$&
$\begin{pmatrix}5644.9\\5561.2\\5507.9\\5364.9\end{pmatrix}$\\
\bottomrule[1.0pt]
\multicolumn{4}{l|}{$nncc\bar {b}(I=1)$}&\multicolumn{3}{l}{$nncc\bar {b}(I=0)$}\\
$J^p$&Eigenvalue&$(\Sigma_{c}B_c)$&$(\Xi_{cc}B)$&Eigenvalue&$(\Sigma_{c}B_c)$&$(\Xi_{cc}B)$\\
\bottomrule[0.7pt]
$\frac52^-$&82.9&8858.4&9051.1&
20.5&8796.0&8988.7\\
$\frac{3}{2}^{-}$ &
$\begin{pmatrix}105.1\\84.1\\36.1\\13.6\end{pmatrix}$ &
$\begin{pmatrix}8880.5\\8859.5\\8811.6\\8789.1\end{pmatrix}$&
$\begin{pmatrix}9073.2\\9052.2\\9004.3\\8981.8\end{pmatrix}$&
$\begin{pmatrix}11.5\\-45.4\\-136.7\end{pmatrix}$&
$\begin{pmatrix}8787.0\\8730.0\\8638.8\end{pmatrix}$&
$\begin{pmatrix}8979.7\\8922.7\\8831.5\end{pmatrix}$\\
$\frac{1}{2}^{-}$ &
$\begin{pmatrix}146.6\\75.8\\27.3\\-21.3\end{pmatrix}$&
$\begin{pmatrix}8922.1\\8851.3\\8802.8\\8754.1\end{pmatrix}$&
$\begin{pmatrix}9114.8\\9044.0\\8995.5\\8946.8\end{pmatrix}$&
$\begin{pmatrix}-19.4\\-71.0\\-121.9\\-198.8\end{pmatrix}$&
$\begin{pmatrix}8756.1\\8704.4\\8653.6\\8576.7\end{pmatrix}$&
$\begin{pmatrix}8948.9\\8897.2\\8846.3\\8769.4\end{pmatrix}$\\
\bottomrule[0.5pt]
\bottomrule[1.5pt]
\end{tabular}
}
\end{table}

Substituting the parameters into the CMI matrices and diagonalizing them, the pentaquark masses are obtained.
Here, we present the masses with corresponding reference systems for the $nncc\bar{Q}$ states in Table \ref{mass-ssccQ}.
As for the $sscc\bar{Q}$, $nnbb\bar{Q}$, and $ssbb\bar{Q}$ states,
we only present the obtained eigenvalues in Tables \ref{eigenvalue1} and \ref{eigenvalue3} and the values of $M_{ref}-\langle H_{\rm CMI}\rangle_{ref}$ in Table \ref{reference} in Appendix. For their mass spectrum, we could obtain the values following the Eq. \ref{Eq11}.
In these systems, a state is explicitly exotic if the flavor of $Q$ is different from the heavy quarks.

For the $nncc\bar{c}$ states, there are two types of reference systems we can adopt, $(c\bar{c})(cnn)$ and $(\bar{c}n)(ccn)$. The mass $M_{\Xi_{cc}} =3621.4$ MeV measured by the LHCb Collaboration is used in the latter case. We assume that the spin of ${\Xi_{cc}}$ is $1/2$ although it has not been determined yet. If the spin is $3/2$, the pentaquark masses estimated with the threshold relating to $M_{\Xi_{cc}}$ would be shifted downward by 64 MeV according to Ref. \cite{Li:2018vhp}, but the gaps are the same. As for the $nnbb\bar{Q}$, $sscc\bar{Q}$, and $ssbb\bar{Q}$ systems, we can similarly adopt two types of reference systems, $(Q\bar{Q})(Qqq)$ and $(\bar{Q}q)(QQq)$. Because other doubly heavy baryons except the $\Xi_{cc}$ are not observed yet in experiments, the theoretical values $\Omega_{cc}=3730.4$ MeV, $\Xi_{bb}=10093.0$ MeV, and $\Omega_{bb}=10193.0$ MeV in Table \ref{comp} are used. In Ref. \cite{Yin:2019bxe}, the similar theoretical values of $\Omega_{cc}$, $\Xi_{bb}$, and $\Omega_{bb}$ are obtained by a confining, symmetry-preserving regularization of a $vector\times vector$ contact interaction.

From Table \ref{mass-ssccQ}, it is obvious that the pentaquark masses will change when one adopts different reference systems, which indicates that the estimation method with Eq. \eqref{Eq11} should be further improved. If the adopted model can reproduce all the hadron masses accurately, different reference thresholds should lead to the same result.

Table \ref{mass-ssccQ} shows us that the obtained $nncc\bar{c}$ ($nncc\bar{b}$) masses with the reference threshold $\Sigma_{c}J/\psi$ ($\Sigma_{c}B_{c}$) are lower than those with $\Xi_{cc}\bar{D}$ ($\Xi_{cc}B$). This feature is consistent with our anticipation since $\Delta_{\Sigma_{c}}+\Delta_{J/\psi} > \Delta_{\Xi_{cc}}+\Delta_{D}$ and $\Delta_{\Sigma_{c}}+\Delta_{B_{c}} > \Delta_{\Xi_{cc}}+\Delta_{B}$ from Table \ref{theoreticalvalue}. At present, it is not clear which type threshold gives more reasonable masses. For a pentaquark, the effective attraction is probably not strong and maybe a higher mass would be more reasonable \cite{Zhou:2018pcv}. However, the choice of reference scale does not affect the mass splittings.

\begin{figure*}[htbp]
\begin{tabular}{ccc}
\includegraphics[width=235pt]{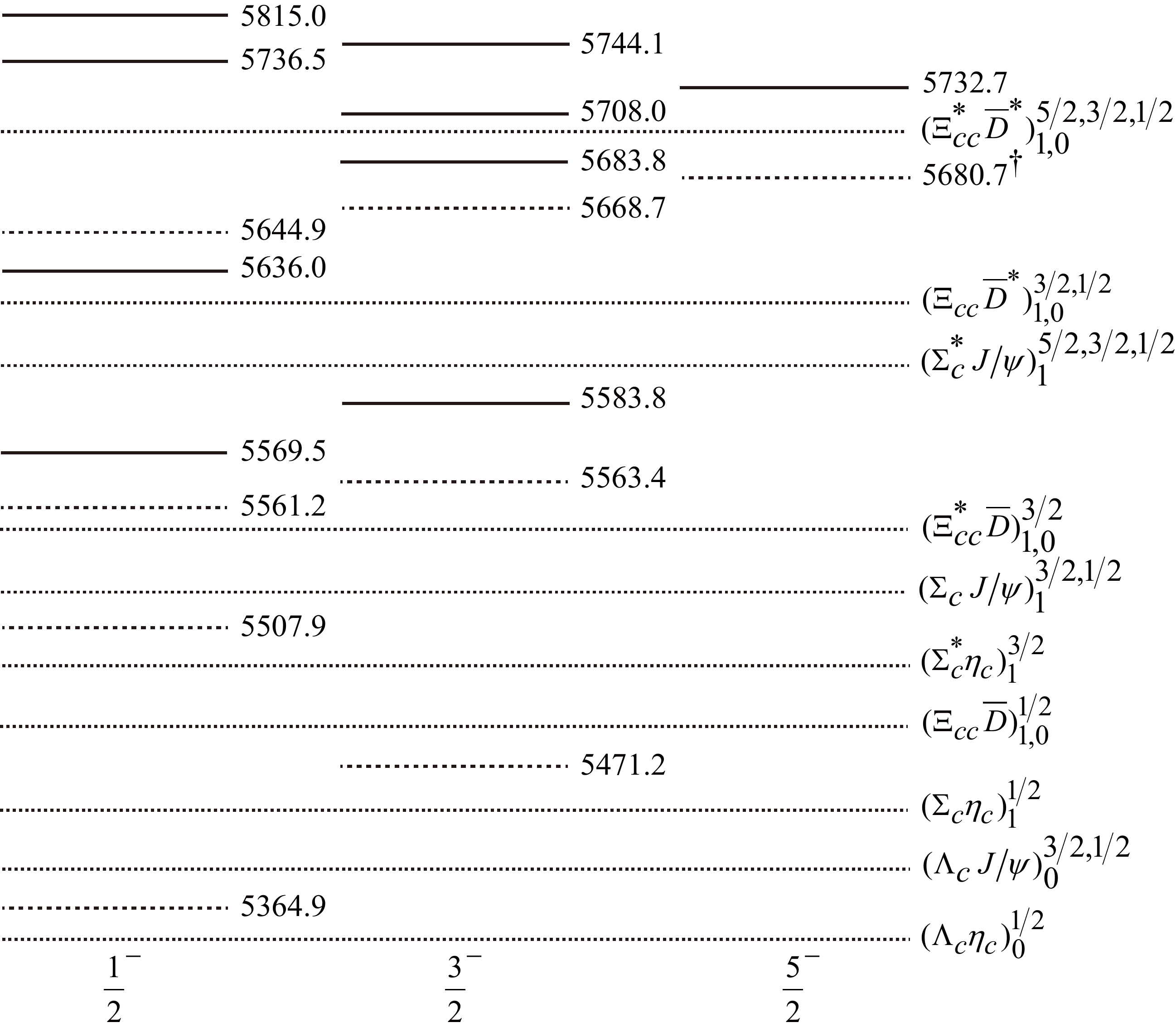}&$\qquad$&\includegraphics[width=235pt]{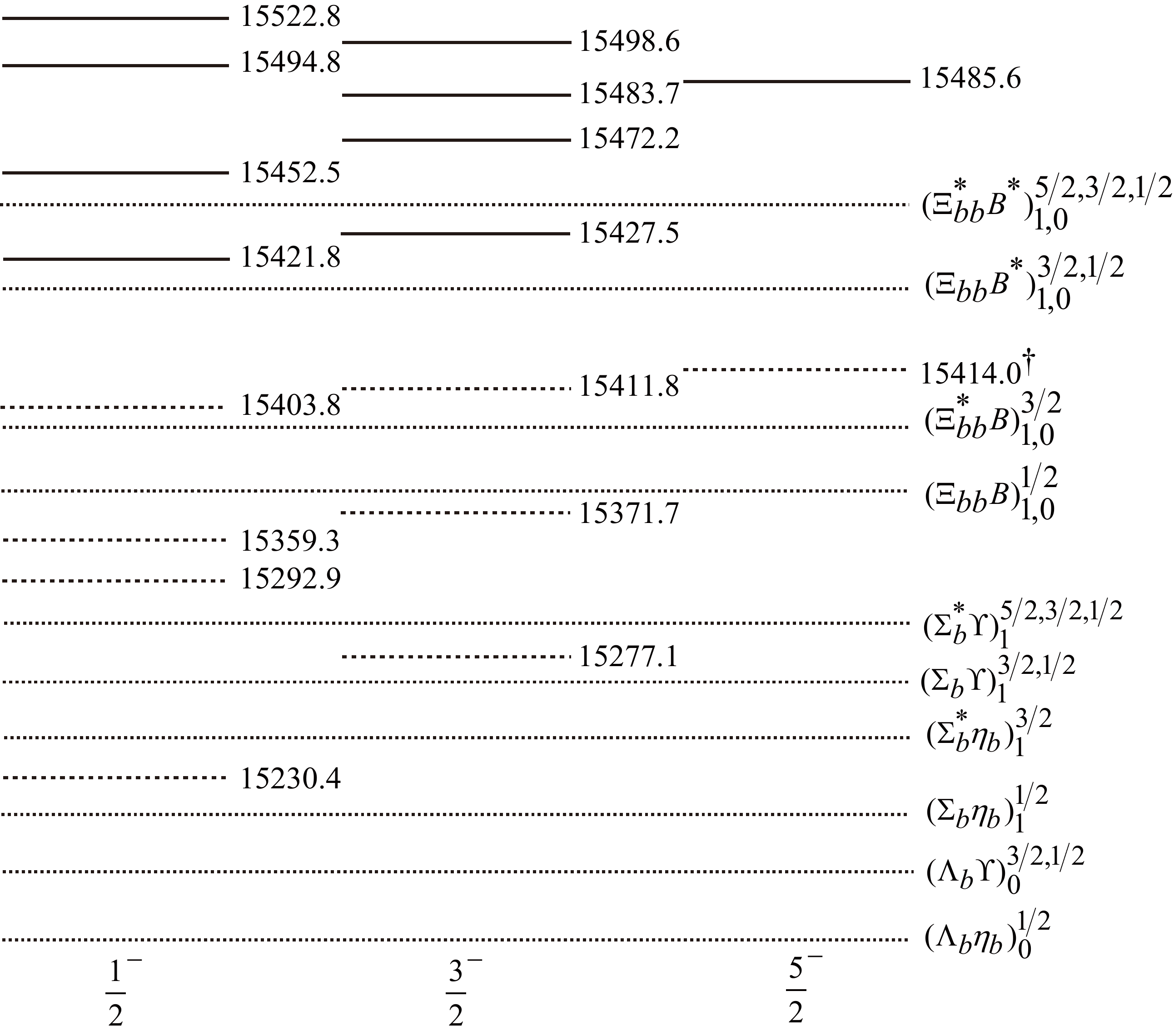}\\
(a) \begin{tabular}{c} $I=1$ (solid) and ($I=0$) (dashed) $nncc\bar{c}$ states\end{tabular} &&(b) $I=1$ (solid) and $I=0$ (dashed) $nnbb\bar{b}$ states\\
&&\\
\includegraphics[width=235pt]{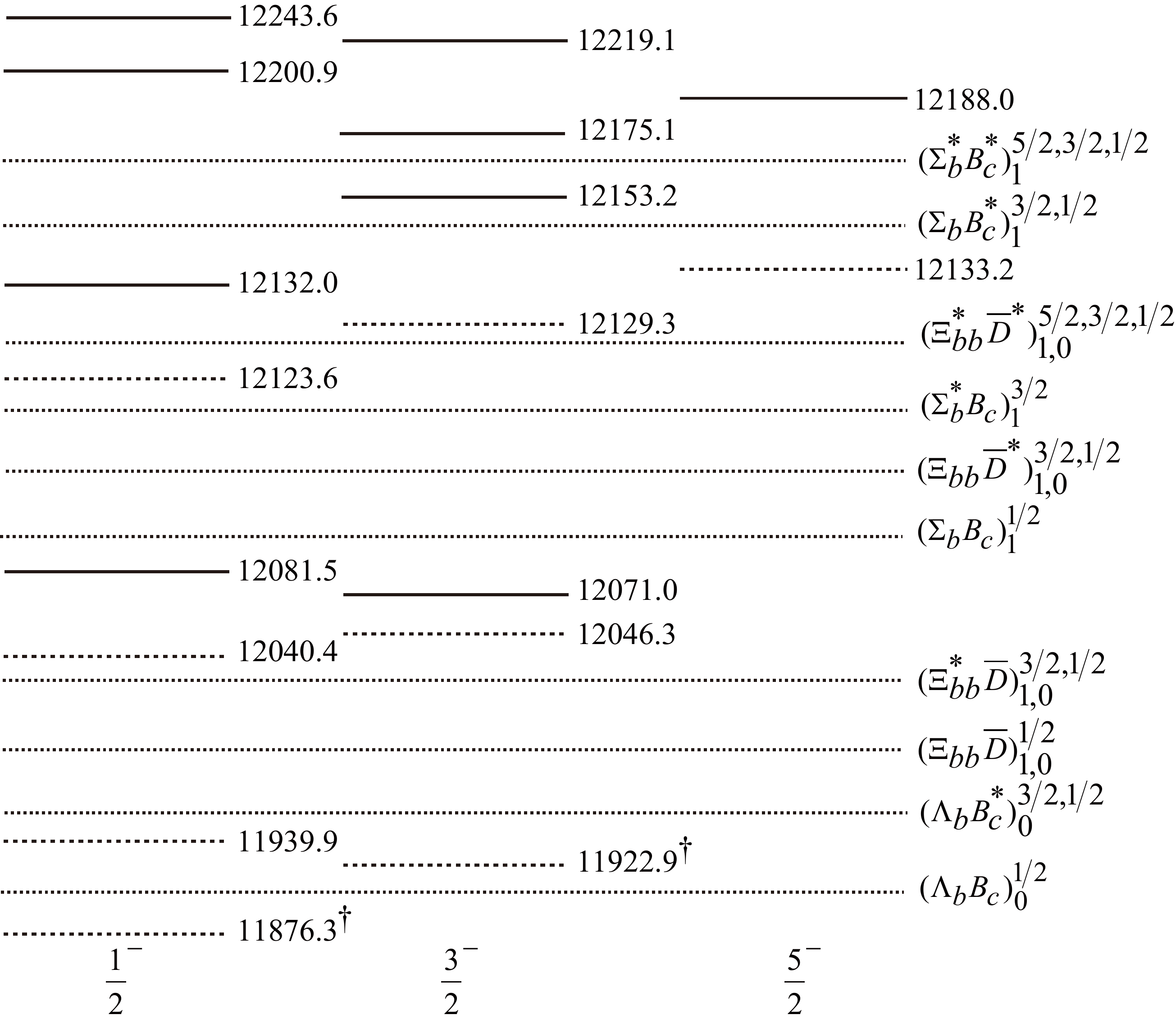}&$\qquad$&\includegraphics[width=235pt]{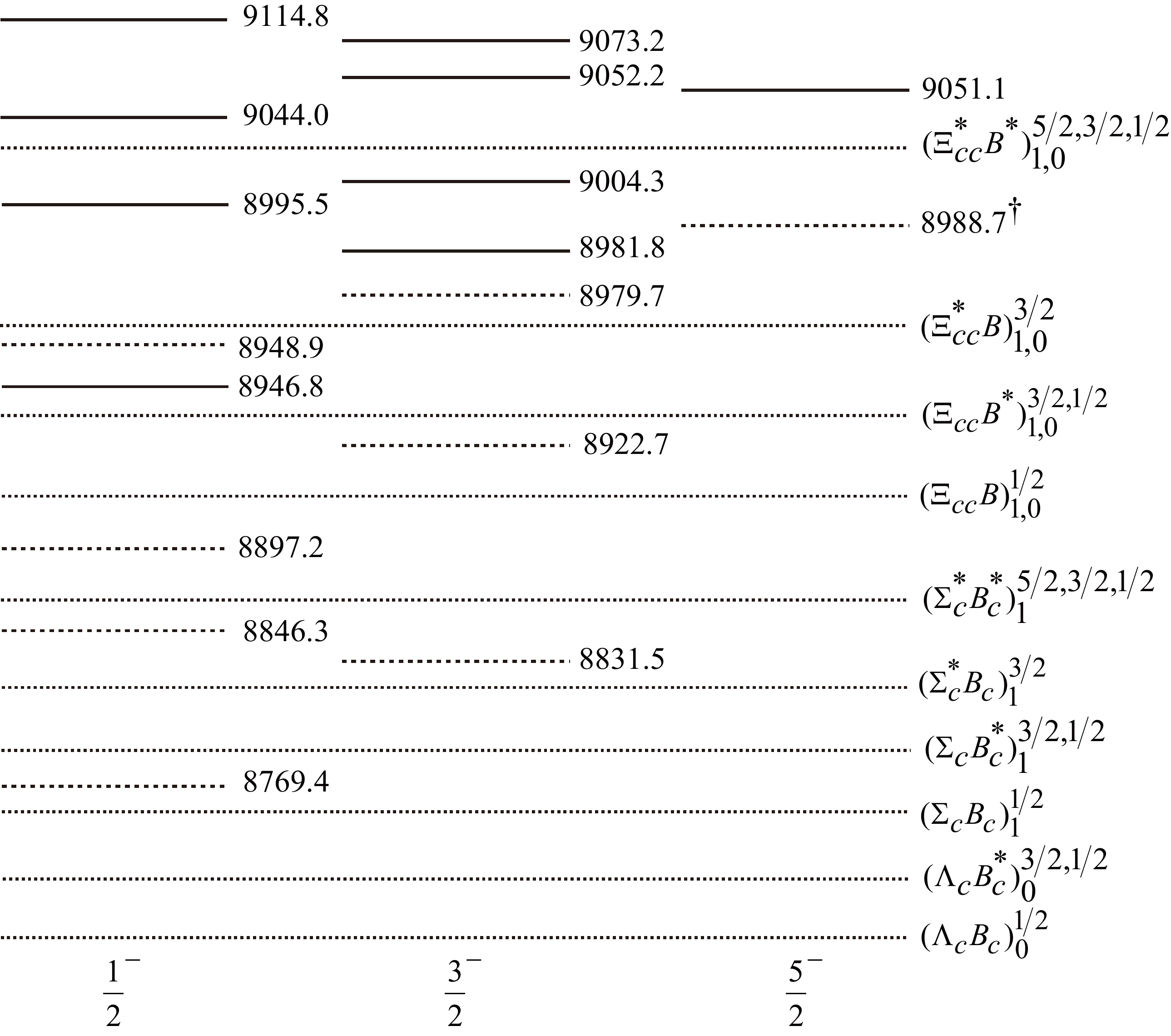}\\
(c) \begin{tabular}{c}($I=1$) (solid) and ($I=0$) (dashed) $nnbb\bar{c}$ states \end{tabular}&&(d) $I=1$ (solid) and $I=0$ (dashed) $nncc\bar{b}$ states\\
\end{tabular}
\caption{Relative positions (units: MeV) for the $nncc\bar{c}$, $nnbb\bar{b}$, $nnbb\bar{c}$ and $nncc\bar{b}$ pentaquark states.
The dotted lines indicate various baryon-meson thresholds. When the isospin (spin) of an initial pentaquark state is equal to a number in the subscript (superscript) of a baryon-meson state, its decay into that baryon-meson channel through S-wave is allowed by the isospin (angular momentum) conservation.
We have adopted the masses estimated with the reference thresholds of (a) $\Xi_{cc}\bar{D}$, (b) $\Xi_{bb}B$, (c) $\Sigma_{b}B^-_c$ and (d) $\Xi_{cc}B$.
The relatively stable states judged with the observed hadrons have been marked with a dagger.
}\label{fig-nnccQ}
\end{figure*}
\begin{figure*}[htbp]
\begin{tabular}{ccc}
\includegraphics[width=235pt]{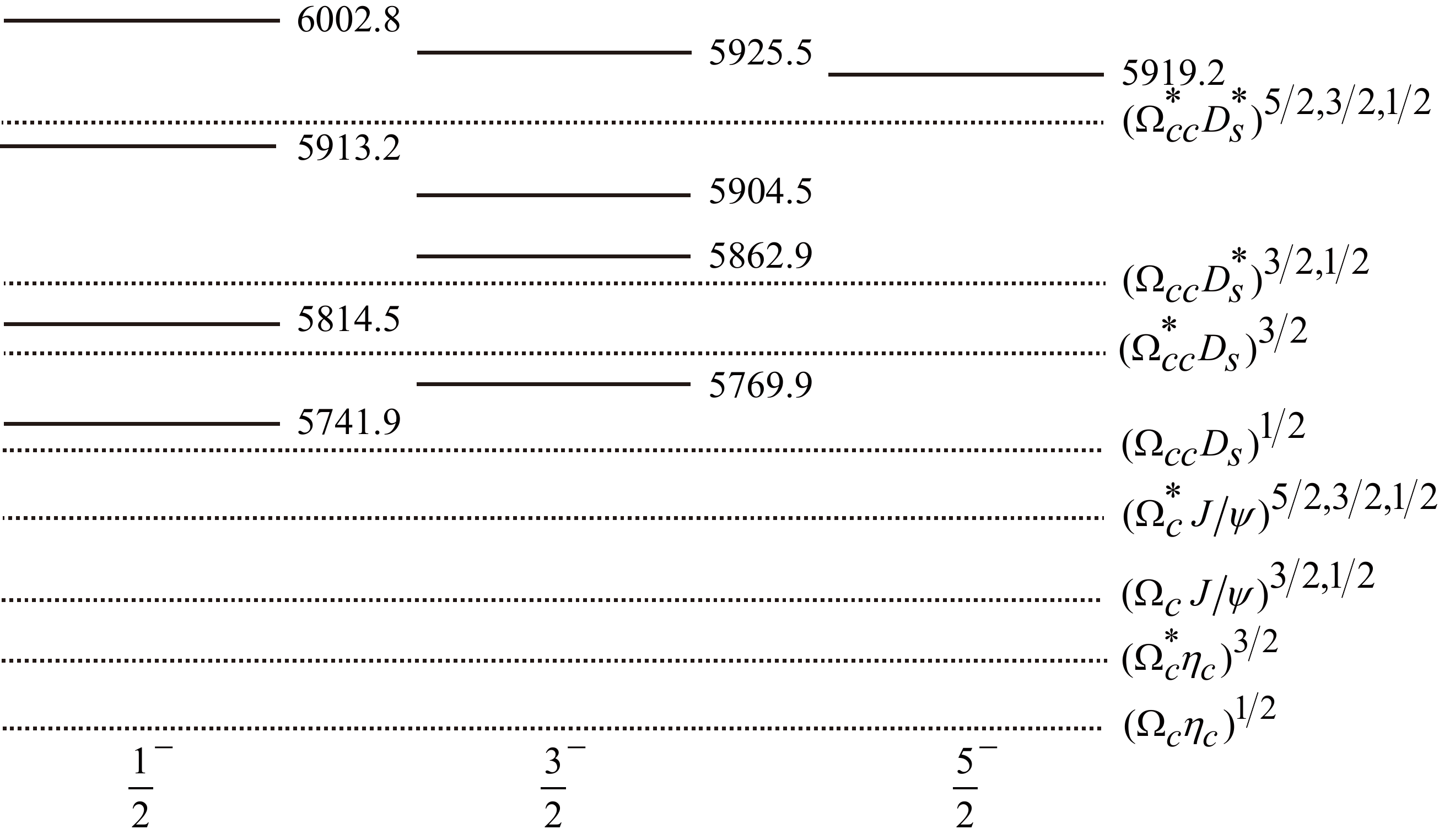}&$\qquad$&\includegraphics[width=235pt]{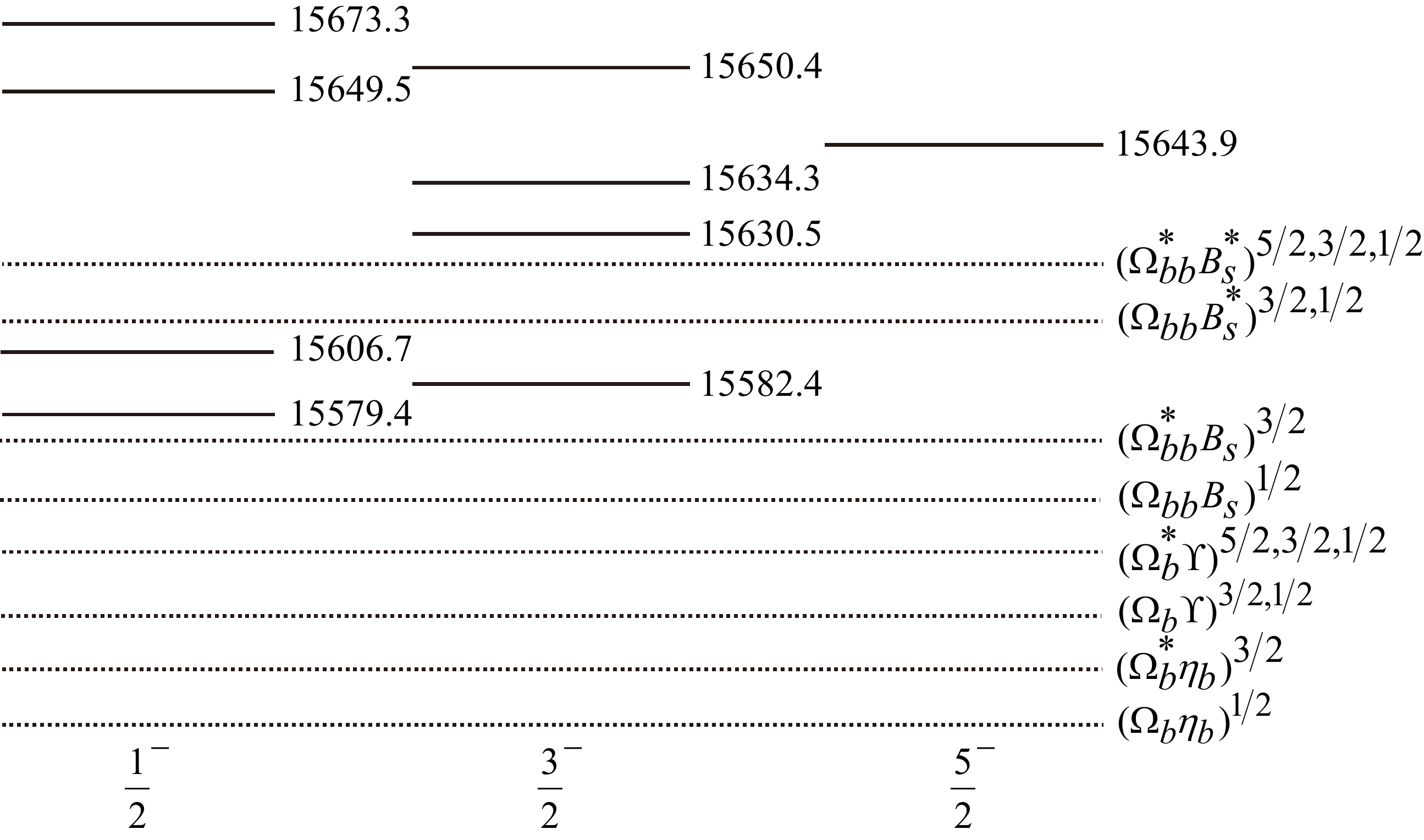}\\
(a) $sscc\bar{c}$ states &&(b) $ssbb\bar{b}$ states\\
&&\\
\includegraphics[width=235pt]{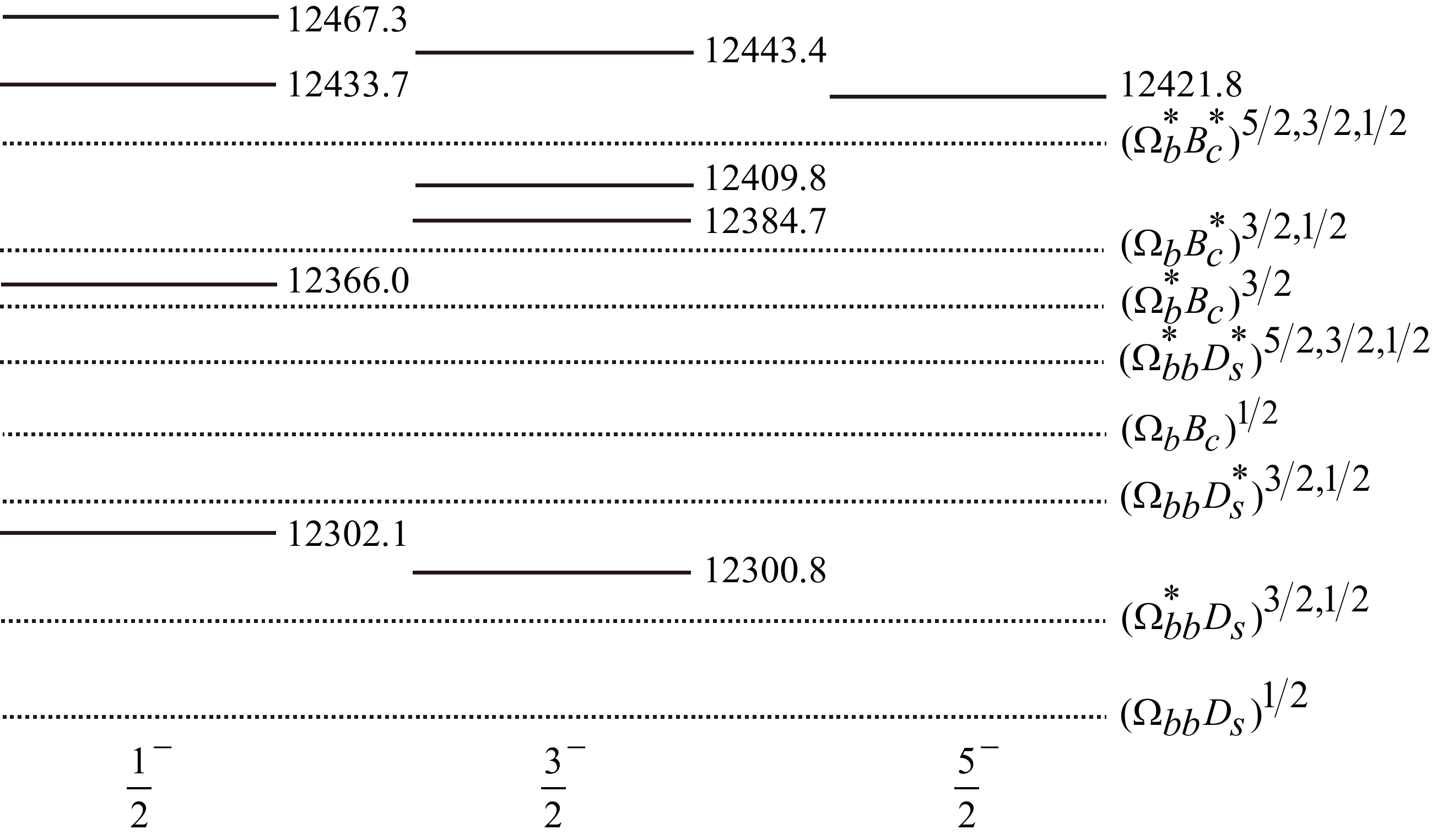}&$\qquad$&\includegraphics[width=235pt]{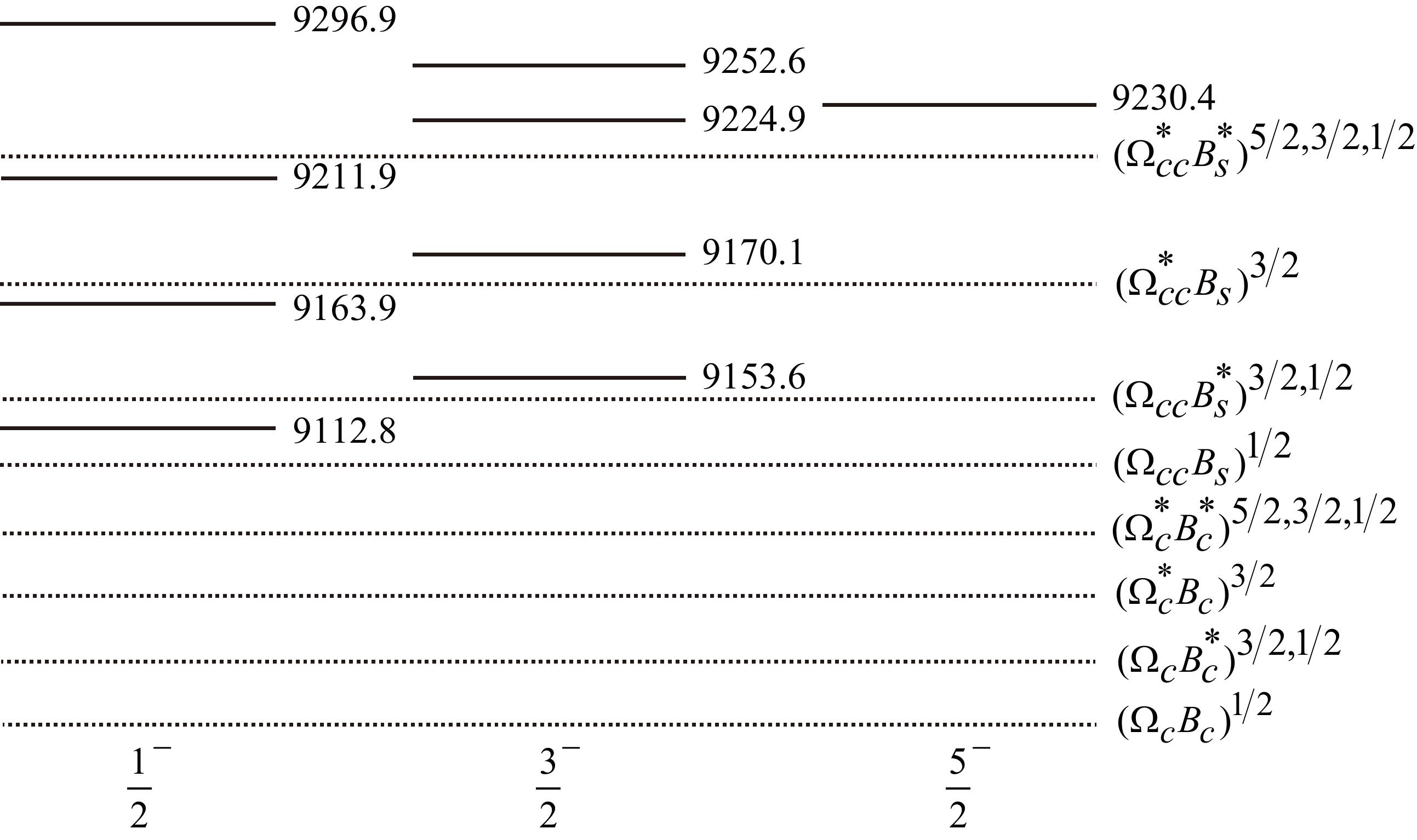}\\
(c) $ssbb\bar{c}$ states &&(d)  $sscc\bar{b}$ states\\
\end{tabular}
\caption{Relative positions (units: MeV) for the $sscc\bar{c}$, $ssbb\bar{b}$, $ssbb\bar{c}$, and $sscc\bar{b}$ pentaquark states. The dotted lines indicate various baryon-meson thresholds. When the spin of an initial pentaquark state is equal to a number in the superscript of a baryon-meson state, its decay into that baryon-meson channel through S-wave is allowed by the angular momentum conservation. We have adopted the masses estimated with the reference thresholds of (a) $\Omega_{cc}D_s^-$, (b) $\Omega_{bb}B_s^0$, (c) $\Omega_{b}B_c^-$, and (d) $\Omega_{cc}B_s^0$.
}\label{fig-ssccQ}
\end{figure*}

In showing the spectra in the figure form, we use the higher pentaquark masses although relevant estimations rely on the masses of the not-yet-observed $QQq$ states. The diagrams of Figs. \ref{fig-nnccQ} and \ref{fig-ssccQ}  illustrate relative positions of the $nncc\bar{c}$, $nnbb\bar{b}$, $nnbb\bar{c}$, $nncc\bar{b}$, $sscc\bar{c}$, $ssbb\bar{b}$, $ssbb\bar{c}$, and $sscc\bar{b}$ states in order. The selected masses are obtained with the reference systems $\Xi_{cc}\bar{D}$, $\Xi_{bb}B$, $\Sigma_{b}B^-_c$, $\Xi_{cc}B$,
$\Omega_{cc}D_s^-$, $\Omega_{bb}B_s^0$, $\Omega_{b}B_c^-$, and $\Omega_{cc}B_s^0$, respectively. The thresholds for relevant rearrangement decay patterns are also displayed.

For the $nncc\bar{c}$ system, the $I=0$ states have generally lower masses than the $I=1$ states.
The quantum numbers for both the lowest and the highest states are $J^P={1/2}^-$.
From the diagrams (b), (c), and (d) of Fig. \ref{fig-nnccQ}, one sees similar features for the $nnbb\bar{b}$, $nnbb\bar{c}$, and $nncc\bar{b}$ systems.

As for the stability of the pentaquark states, their dominant decay modes should be related with the rearrangement mechanism. Now we move on to such decays. One has to consider the constraints from the angular momentum conservation, isospin conservation, parity conservation, and so on when discussing allowed decay channels. For convenience, we have marked the spin and isospin of the baryon-meson channels in the superscripts and subscripts of their symbols in Fig. \ref{fig-nnccQ}, respectively.
For the $sscc\bar{Q}$ and $ssbb\bar{Q}$ states, only one isospin is possible and no label is given explicitly. Of course, whether the decay can happen or not is also kinematically constrained by the pentaquark mass which depends on models. In the following discussions, we assume that the obtained masses shown in the figures are all reasonable.

For the $nncc\bar{c}$ states, they look like excited $nnc$ baryons. Because only orbital or radial excitation energy cannot explain their high masses, the states once observed are good candidates of compact $nncc\bar{c}$ pentaquark states or hadronic molecules.
To distinguish these two configurations, decay properties would be helpful.
We here just discuss relevant rearrangement decay patterns.
In the case of $I(J^P)=1({5/2}^-)$, the possible S-wave decay channels are $\Sigma_{c}^{*}J/\psi$ and $\Xi_{cc}^*\bar{D}^*$.
In the case of $I(J^P)=0({5/2}^-)$, the possible S-wave decay channel is only $\Xi_{cc}^*\bar{D}^*$.
The $I(J^P)=0({5/2}^-)$ isoscalar pentaquark is a candidate of stable state.
We mark it in Fig. \ref{fig-nnccQ}(a) with a dagger.
In the case of $I(J^P)=1({3/2}^-)$, the possible S-wave channels are $\Xi^{*}_{cc}{\bar{D}}^{*}$, $\Xi_{cc}{\bar{D}}^{*}$, $\Sigma^{*}_{c}J/\psi$, $\Xi^{*}_{cc}{\bar{D}}$, $\Sigma_{c}J/\psi$, and $\Sigma^{*}_{c}\eta_{c}$.
In the case of $I(J^P)=0({3/2}^-)$, the possible S-wave channels are $\Xi^{*}_{cc}{\bar{D}}^{*}$, $\Xi_{cc}{\bar{D}}^{*}$ and $\Lambda_{c}J/\psi$.
In the case of $I(J^P)=1({1/2}^-)$, the possible S-wave channels are $\Xi^{*}_{cc}{\bar{D}}^{*}$, $\Xi_{cc}{\bar{D}}^{*}$, $\Sigma_{c}^{*}J/\psi$, $\Sigma_{c}J/\psi$, $\Xi_{cc}{\bar{D}}$, and $\Sigma_{c}\eta_{c}$.
In the case of $I(J^P)=0({1/2}^-)$, the possible S-wave channels are $\Xi^{*}_{cc}{\bar{D}}^{*}$, $\Xi_{cc}{\bar{D}}^{*}$, $\Xi_{cc}{\bar{D}}$, $\Xi^{*}_{cc}{\bar{D}}$, $\Lambda_{c}J/\psi$, and $\Lambda_{c}\eta_{c}$.
The observation of any one of the mentioned decay patterns could provide hints for the existence of a $nncc\bar{c}$ pentaquark state.
Because the lowest $I(J^P)=0(1/2)^{-}$ state is much lower than the $\Xi_{cc}\bar{D}$ threshold, if an observed state in $\Lambda_c\eta_c$ (or $\Lambda_cJ/\psi$) is around 5.4 GeV, this state would be more likely to be a compact pentaquark than a $\Xi_{cc}\bar{D}$ molecule.
If the spin of the observed $\Xi_{cc}$ by LHCb is $3/2$, $\Xi_{cc}\to\Xi_{cc}^*$ and the estimated pentaquark masses will be reduced by 64 MeV. The stability of the pentaquark states is not affected.

For the $nnbb\bar{c}$ states shown in Fig. \ref{fig-nnccQ}(c), they are explicitly exotic.
Since the $nbb$ states and the excited $B_c^*$ have not yet been observed in experiments, we use the theoretical masses of $B_c^*$, $\Xi_{bb}$, and $\Xi_{bb}^*$ in Table \ref{comp} to check the pentaquark stability. Now, it is easy to see that the lowest-lying states with $I(J^P)=0({1/2}^{-})$ and $I(J^P)=0({3/2}^{-})$ are both stable. The situation for the $nncc\bar{b}$ ($nnbb\bar{b}$) states can be analyzed similar to the $nncc\bar{c}$ ($nnbb\bar{c}$) case, but now all of them are explicitly (implicitly) exotic.

For the $sscc\bar{c}$, $ssbb\bar{b}$, $sscc\bar{b}$, and $ssbb\bar{c}$ states, their properties are similar to those of $nncc\bar{Q}$ $(I=1)$ and $nnbb\bar{Q}$ $(I=1)$. Here, we also use the theoretical masses of $B^*_{c}$, $\Omega_{cc}$, $\Omega^{*}_{cc}$, $\Omega_{bb}$, and $\Omega^{*}_{bb}$ to discuss the possible decay channels. In the $sscc\bar{c}$ case shown in Fig. \ref{fig-ssccQ}(a), any possible pentaquark is above their allowed rearrangement decay channels and thus there is no stable state. One does not find stable states in the $ssbb\bar{b}$ and $sscc\bar{b}$ cases, either. In the $ssbb\bar{c}$ system, the lowest-lying $(J^P)=({3/2}^{-})$ pentaquark is slightly above its decay channel $\Omega^{*}_{bb}D_{s}^-$. Probably it is not a broad state.


\subsection{The $nnbc\bar{Q}$ and $sscb\bar{Q}$ pentaquark states}

All these $nnbc\bar{Q}$ and $ssbc\bar{Q}$ states are implicitly exotic. To estimate their masses, we can use three types of reference systems, $(qqc)$-$(b\bar{Q})$, $(qqb)$-$(c\bar{Q})$, and $(qbc)$-$(q\bar{Q})$.
We present the obtained eigenvalues in Tables \ref{eigenvalue1} and \ref{eigenvalue2} and the values of $M_{ref}-\langle H_{\rm CMI}\rangle_{ref}$ in Table \ref{reference} in Appendix for the $nnbc\bar{Q}$ and $ssbc\bar{Q}$ states, respectively. 
From relevant calculation, the results with these three types of reference systems are slightly different.

The masses we use are obtained with the reference thresholds of $\Sigma_{b}J/\psi$, $\Xi_{bc}B$, $\Omega_{b}J/\psi$, and $\Omega_{bc}B_s$ channels for the $nnbc\bar{c}$, $nnbc\bar{b}$, $ssbc\bar{c}$, and $ssbc\bar{b}$ states, respectively.
Moreover, 16 rearrangement decay channels are involved for the $nnbc\bar{c}$ and $nncb\bar{b}$ states and 12 channels are involved for the $ssbc\bar{c}$ and $sscb\bar{b}$ states.

We first check possible stable pentaquarks in the $nnbc\bar{c}$ case. The lowest $J^P=1/2^-$ and $J^P=3/2^-$ states both have rearrangement decay channels and should not be very narrow.
On the contrary, the lowest $I(J^P)=0(5/2)^-$ state is below the possible decay channel $\Xi_{cb}^*\bar{D}_{s}^{*}$ and it is considered a relative state. Similarly, the only possible stable $nncb\bar{b}$ pentaquark has the quantum numbers $I(J^P)=0(5/2^-)$ if the mass of $\Xi_{bc}^*$ is larger than 6870 MeV.
Lastly, it seems that there is no stable $ssbc\bar{c}$ or $sscb\bar{b}$ pentaquark state.
The possible stable pentaquarks have been shown in Table \ref{aaa}.

\subsection{The $nscc\bar{Q}$ and $nsbb\bar{Q}$ pentaquark states}



For the $nscc\bar{Q}$ and $nsbb\bar{Q}$ states, there are also three types of reference systems we can use to estimate the masses, $(nsQ)$-$(Q\bar{Q})$, $(nQQ)$-$(s\bar{Q})$, and $(sQQ)$-$(n\bar{Q})$. For example, we estimate the $nscc\bar{c}$ masses with the thresholds of $\Xi_{c}^\prime J/\psi$, $\Xi_{cc}D_s^-$, and $\Omega_{cc}\bar{D}$ channels. Similarly, we use the reference systems of $\Xi_{b}^\prime B_{c}^-$, $\Xi_{bb}D_s^-$, and $\Omega_{bb}\bar{D}$ to estimate the $nsbb\bar{c}$ masses.
The eigenvalues and the values of $M_{ref}-\langle H_{\rm CMI}\rangle_{ref}$ for the $nscc\bar{Q}$ and $nsbb\bar{Q}$ systems are presented in Tables \ref{eigenvalue1} and \ref{reference} of Appendix, respectively.

Of these states, the $nscc\bar{b}$ and $nsbb\bar{c}$ pentaquarks are explicitly exotic. The observation of such a state in future measurements will be an important finding, in particular when the state is narrow.
From 
relevant calculation, the $nscc\bar{c}$ pentaquark masses estimated with the $(nsQ)$-$(Q\bar{Q})$ type reference systems are lower than those with other type thresholds.
Such states are easy to be identified as five-quark states because of their high masses, although they are implicitly exotic.
The situation is different from the $nscc\bar{n}$ or $nscc\bar{s}$ states studied in Ref. \cite{Zhou:2018pcv}. It is not easy to distinguish such a pentaquark state from a 3q baryon once it is observed.

From Table \ref{eigenvalue2} and the rough values of the doubly heavy 3q baryons in Table \ref{comp}, it seems that no stable pentaquark states exist in the $nscc\bar{c}$, $nsbb\bar{b}$, and $nscc\bar{b}$ systems. However, 
the lowest $J^P={1/2}^-$, $J^P={3/2}^-$, and $J^P={5/2}^-$ states are below any possible rearrangement decay channels and they are possibly stable. Of course, if its mass is underestimated, they may also decay into $\Xi_{bb}D_s^-$ (and probably $\Omega_{bb}\bar{D}$), $\Omega^{*}_{bb}\bar{D}$, and $\Omega^{*}_{bb}\bar{D}^{*}$, respectively.

\subsection{The $nsbc\bar{Q}$ pentaquark states}

\begin{figure*}[htbp]
\begin{tabular}{ccc}
\includegraphics[width=240pt]{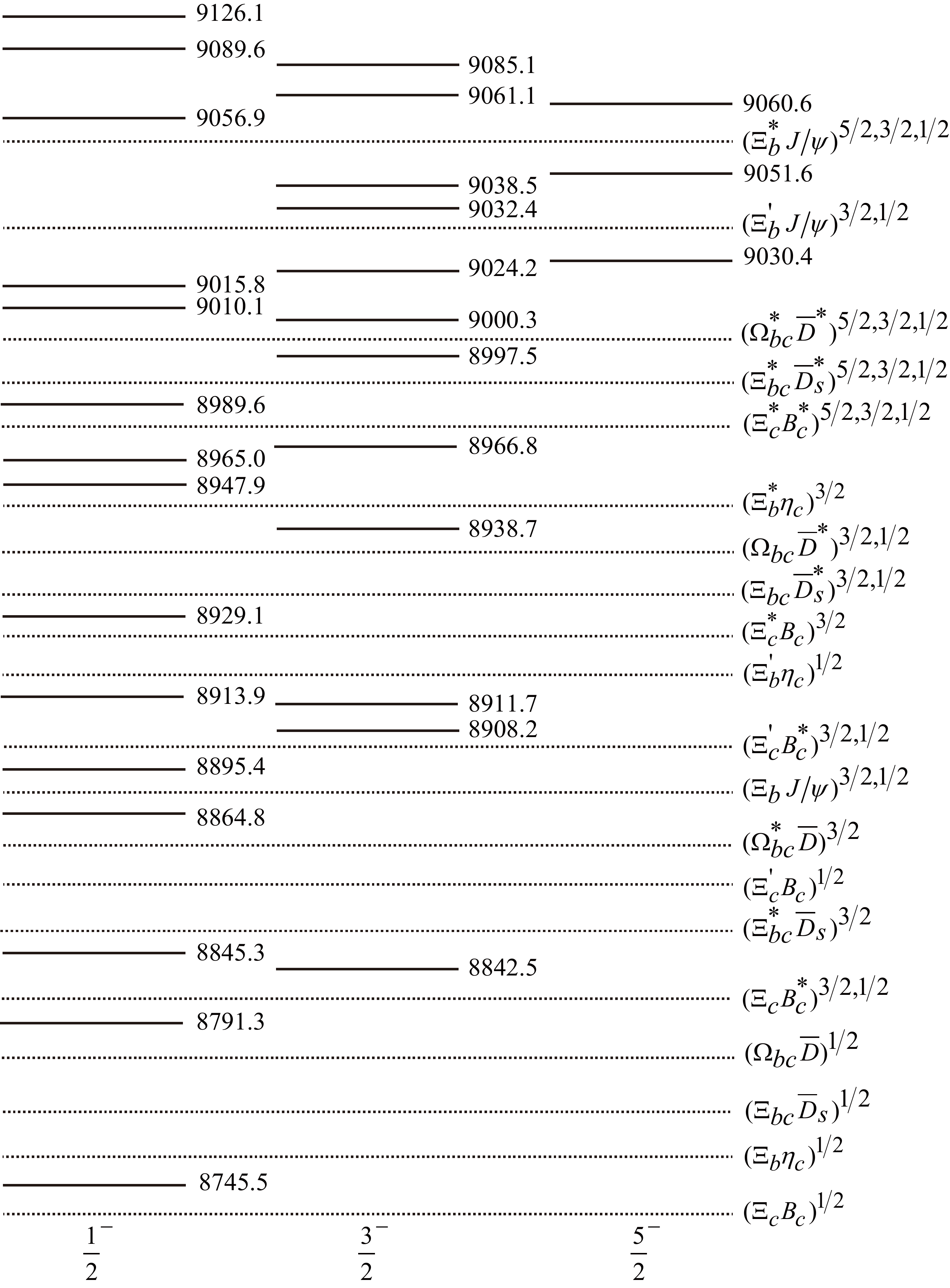}&$\qquad$&
\includegraphics[width=240pt]{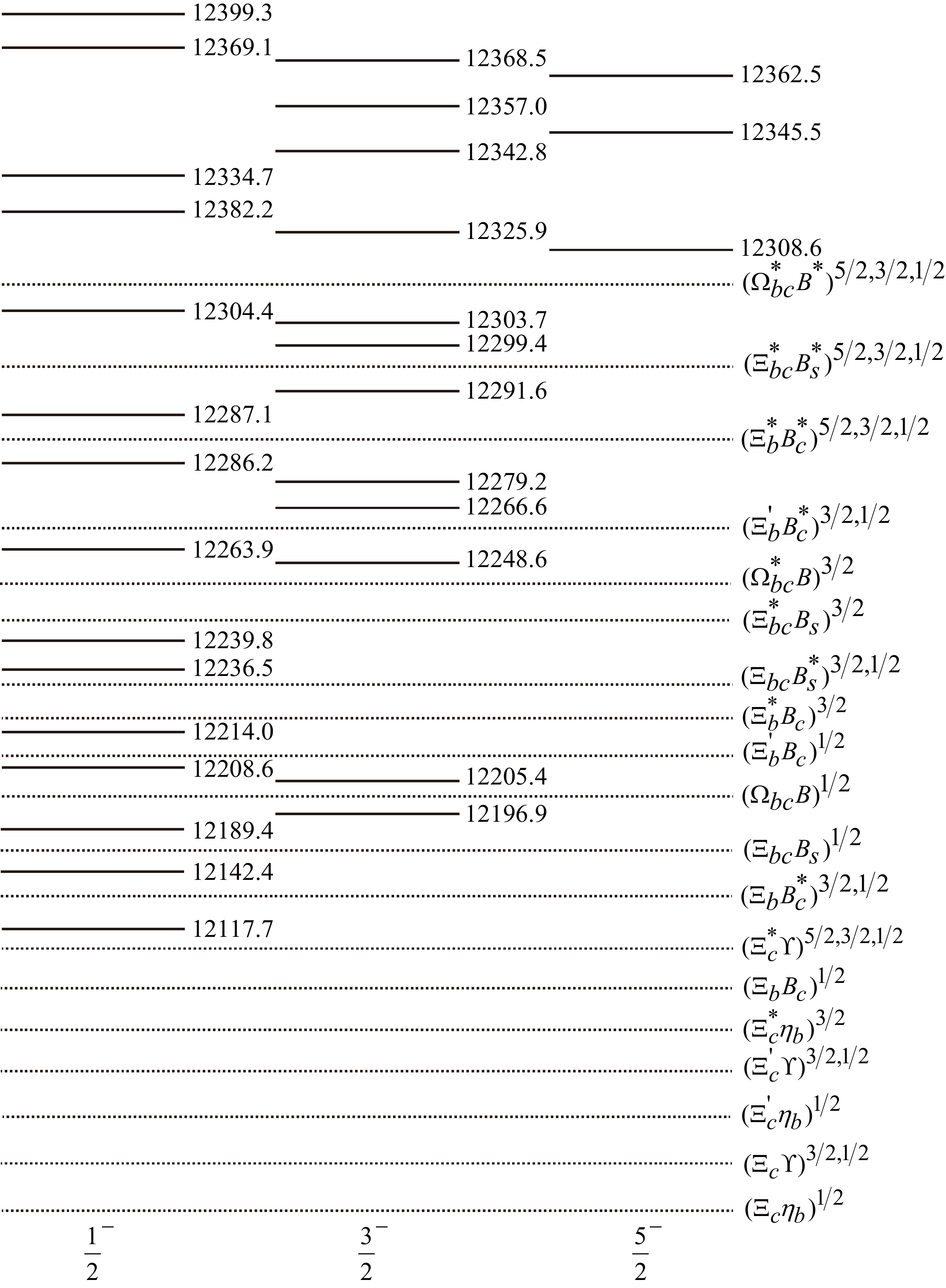}\\
(a) \begin{tabular}{c}  $nsbc\bar{c}$ states\end{tabular} &&(b)  $nscb\bar{b}$ states\\
&&\\
\end{tabular}
\caption{Relative positions (units: MeV) for the $nsbc\bar{c}$ and $nsbc\bar{b}$ pentaquark states. The dotted lines
indicate various baryon-meson thresholds. When the spin of an initial pentaquark state is
equal to a number in the superscript of a baryon-meson state, its
decay into that baryon-meson channel through S-wave is allowed
by the angular momentum conservation. We have adopted
the masses estimated with the reference thresholds of (a) $\Xi_{b}^{'}J/\psi$ and (b) $\Omega_{bc}B$.}\label{fig-nsbcQ}
\end{figure*}

The $nsbc\bar{Q}$ states are implicitly exotic. Their wave functions are not constrained from the Pauli principle. The number of wave function bases for a pentaquark with given quantum numbers is bigger than that for other states. After diagonalizing the Hamiltonian, one gets numbers of possible pentaquark states. To estimate the $nsbc\bar{c}$ ($nscb\bar{b}$) masses, we use four different types of reference systems, $\Xi_{c}^\prime B_{c}^-$ ($\Xi_c\Upsilon$), $\Xi_{b}^\prime J/\psi$ ($\Xi_b^\prime B_{c}^+$), $\Xi_{bc}D_{s}^-$ ($\Xi_{bc}B_{s}^0$), and $\Omega_{bc}\bar{D}$ ($\Omega_{bc}B$).
We present the obtained eigenvalues and the values of $M_{ref}-\langle H_{\rm CMI}\rangle_{ref}$ in Tables \ref{eigenvalue2} and \ref{reference} in Appendix for the $nsbc\bar{Q}$ states, respectively.
Meanwhile, the results with these four types of reference systems are slightly different and we use the highest values to plot the Fig. \ref{fig-nsbcQ}.

The spectra for the $nsbc\bar{Q}$ pentaquark states with the $\Xi_{b}^\prime J/\psi$ or $\Omega_{bc}B$ threshold are shown in Fig. \ref{fig-nsbcQ}. From the figure and the masses given in Table \ref{comp}, it is difficult to find stable pentaquarks in these systems. Only the lowest $nsbc\bar{c}$ pentaquark is slightly above the $\Xi_{c}B_{c}^-$ threshold and is possibly a state without broad width. Of course, the $nsbc\bar{c}$ states can be searched for in the $\Xi_cB_c^-$ or $\Xi_b\eta_c$ channel in future experiments. If such a state could be observed, its exotic nature can be easily identified, a situation different from the $nsbc\bar{q}$ case \cite{Zhou:2018pcv}.

\section{Discussions and summary}\label{sec5}

Recently, the observation of the $P_c (4312)$, $P_c (4440)$ and $P_c (4457)$ at LHCb \cite{Aaij:2019vzc} gave us significant evidence for the existence of pentaquak states, which motivates us to study the ground compact $qqQQ\bar{Q}$ ($q = u, d, s$ and $Q = b, c$) pentaquark states within the CMI model. In the considered pentaquark systems, the $qqbb\bar{c}$, $qqcc\bar{b}$ states are explicitly exotic and are easy to be identified. Other states can also be easily identified as exotic baryons because their large masses could not be understood without an excited $Q\bar{Q}$ pair.

In this work, we have firstly constructed the flavor-color-spin wave functions for the $qqQQ\bar{Q}$ pentaquark states from the SU(3) and SU(2) symmetries and Pauli Principle. We extract the effective coupling constants from the mass splittings between conventional hadrons. Based on these, we systematically calculate the color-magnetic interaction for these pentaquark states and obtain the corresponding mass gaps. Then, various reference thresholds are used to estimate the masses of these states. Some theoretical results for the masses of the doubly heavy 3q baryons are adopted in our estimation. At last, we analyze the stability and possible rearrangement decay channels of the $qqQQ\bar{Q}$ pentaquark states.

We have shown the mass spectra and rearrangement decay patterns in the figure form. Following Figs. \ref{fig-nnccQ}, \ref{fig-ssccQ} and Tables \ref{eigenvalue1}, \ref{eigenvalue2} and \ref{eigenvalue3}, we can see ten stable states are possible which are also collected in Table. \ref{aaa}. However, not all of them are really stable states. The reason is that the predicted pentaquarks in the current model may have mass deviations from the case they should be.

\begin{table}[htbp]
\caption{The stable states in the $qqQQ\bar{Q}$ systems from relevant calculations}\label{aaa}
\begin{tabular}{|ccc|ccc|}
\bottomrule[1.5pt]
\bottomrule[0.5pt]
Systems&$I(J^{P})$&Masses&Systems&$I(J^{P})$&Masses\\
\bottomrule[0.7pt]
\multirow{1}*{$nncc\bar{c}$}&$0(5/2^{-})$&5680.7&\multirow{2}*{$nnbb\bar{c}$}&$0(3/2^{-})$&11922.9\\ \cline{1-3}
\multirow{1}*{$nnbb\bar{b}$}&$0(5/2^{-})$&15414.0&&$0(1/2^{-})$&11876.3\\\cline{1-6}
\multirow{1}*{$nncc\bar{b}$}&$0(5/2^{-})$&8988.7&\multirow{3}*{$nsbb\bar{c}$}&$1/2(5/2^{-})$&12214.8\\  \cline{1-3}
\multirow{1}*{$nnbc\bar{c}$}&$0(5/2^{-})$&8882.1&&$1/2(3/2^{-})$&12026.3\\  \cline{1-3}
\multirow{1}*{$nncb\bar{b}$}&$0(5/2^{-})$&12164.2&&$1/2(1/2^{-})$&11983.6\\  \cline{1-3}
\bottomrule[0.5pt]
\midrule[1.5pt]
\end{tabular}
\end{table}

As a general feature, the high spin $J^P=(5/2)^{-}$ pentaquark states should be usually narrow since they have many $D$-wave decay modes but one or two $S$-wave decay modes. This feature is similar to the $\bar{Q}Qqqq$ and $QQqq\bar{q}$ cases \cite{Wu:2017weo,Zhou:2018pcv}. Now, the narrow $\bar{Q}Qqqq$ pentaquark states have been observed. It is also worthwhile to search for the exotic narrow $qqQQ\bar{Q}$ pentaquark states in future experiments.

In the study of the pentaquark states, the number of color-spin structures may be more than ten. The mixing or channel-coupling effects could be important. Such effects should be carefully considered in detail in further studies.

In short summary, we have systematically studied the exotic states with the structure $qqQQ\bar{Q}$ ($q=u,d,s$ and $Q=c,b$). They can be easily identified as explicitly exotic or implicitly exotic pentaquark states once observed. We hope the present study may stimulate further investigations about properties of the $qqQQ\bar{Q}$ pentaquark states from both the theoretical and the experimental aspects.

\section{Acknowledgments}\label{sec6}
This project is supported by the National Natural Science Foundation of China under Grants Nos. 11705072, 11775132 and the China National Funds for Distinguished Young Scientists under Grants No. 11825503.
\section{Appendix}\label{sec6}
In this appendix, we show the CMI Hamiltonian matrix elements with $J^P={5/2}^{-}$, $J^P={3/2}^{-}$, and $J^P={1/2}^{-}$ in Tables \ref{11}, \ref{12}, and \ref{13}, respectively.

Then, we show the eigenvalues of the $nncc(bb)\bar{Q}$, $sscc(bb)\bar{Q}$, $nnbc\bar{Q}$, $ssbc\bar{Q}$, $nscc(bb)\bar{Q}$ and $nsbc\bar{Q}$ systems in Tables \ref{eigenvalue1}, \ref{eigenvalue2} and \ref{eigenvalue3}, respectively.
The corresponding mass $M$ can be obtained with the eigenvalue $\langle H_{\rm CMI} \rangle$
from the following equation
\begin{equation}
M=\langle H_{\rm CMI} \rangle+(M_{ref}-\langle H_{\rm CMI}\rangle_{ref}),
\end{equation}
and we list $M_{ref}-\langle H_{\rm CMI}\rangle_{ref}$ in Table \ref{reference}.

\begin{widetext}
\begin{table*}[htbp]
\centering \caption{The $H_{\rm CMI}$ matrix elements for the case $J^P={\frac52}^{-}$.}\label{11}
\begin{tabular}{|c|ccc|}
\bottomrule[1.0pt]\midrule[0.5pt]
\diagbox{Bases}{Elements}{Bases}&$\phi_{1}\chi_{1}$&$\phi_{2}\chi_{1}$&$\phi_{3}\chi_{1}$\\
\midrule[0.5pt]
$\phi_{1}\chi_{1}$&$\frac{1}{3}[8\alpha+2\zeta+4(\nu+\lambda)]$&$\sqrt2\eta-2\sqrt2\mu$&$-\sqrt2\theta-2\sqrt2\xi$\\
$\phi_{2}\chi_{1}$&$\sqrt2\eta-2\sqrt2\mu$&$\frac{1}{3}(-4\beta+5\zeta+10\lambda-2\nu)$&$-\kappa$\\
$\phi_{3}\chi_{1}$&$-\sqrt2\theta-2\sqrt2\xi$&$-\kappa$&$\frac{1}{3}(4\gamma+5\zeta-2\lambda+10\nu)$\\
\midrule[0.5pt]\bottomrule[1.0pt]
\end{tabular}
\end{table*}

\begin{sidewaystable}[htbp]
\centering \caption{The $H_{\rm CMI}$ matrix elements for the case $J^P={\frac32}^{-}$.}\label{12}
\resizebox{\columnwidth}{!}{
\begin{tabular}{|c|cccccccccccc|}
\bottomrule[1.0pt]\midrule[0.5pt]
\diagbox{Bases}{Elements}{ Bases}&$\phi_{1}\chi_{2}$&$\phi_{1}\chi_{3}$&$\phi_{1}\chi_{4}$&$\phi_{1}\chi_{5}$&$\phi_{2}\chi_{2}$&$\phi_{2}\chi_{3}$&$\phi_{2}\chi_{4}$&$\phi_{2}\chi_{5}$&
$\phi_{3}\chi_{2}$&$\phi_{3}\chi_{3}$&$\phi_{3}\chi_{4}$&$\phi_{3}\chi_{5}$\\
\midrule[0.5pt]
$\phi_{1}\chi_{2}$&$\frac{1}{3}[\begin{smallmatrix}8\alpha+2\zeta\\-6(\nu+\lambda)\end{smallmatrix}]$&$\frac{2\sqrt{5}}{3}(\nu-\lambda)$ &$ \frac{-2\sqrt{10}}{3}\xi $& $\frac{-2\sqrt{10}}{3}\mu$&${\sqrt{2}}(\eta+3\xi)$ &$ -\sqrt{10}\xi$&$2\sqrt{5}\nu$ &$ 0 $ & $\sqrt{2}(\eta+3\mu)$ & $\sqrt{10}\mu $&$ 0 $ & $2\sqrt{5}\lambda$\\
$\phi_{1}\chi_{3}$&$\frac{2\sqrt{5}}{3}(\nu-\lambda)$&$\frac{1}{3}[\begin{smallmatrix}8\alpha-2\zeta+\\2(\nu+\lambda)\end{smallmatrix}]$&$ \frac{2\sqrt{2}}{3}(\theta+\xi)$ & $\frac{-2\sqrt{2}}{3}(\eta+\mu)$ &$\sqrt{10}\xi$ & $-\sqrt{2}(\theta+\xi)$&$2(\xi-\nu)$ & $-2\kappa$ &$\sqrt{10}\mu$ & $-\sqrt{2}(\eta+\mu)$ &$2\kappa$& $-2(\zeta-\lambda)$\\
$\phi_{1}\chi_{4}$&$\frac{-2\sqrt{10}}{3}\xi$ &$\frac{2\sqrt{2}}{3}(\theta+\xi)$ &$\frac{1}{3}(8\epsilon+4\lambda)$ &$\frac23\kappa$ &$2\sqrt{5}\nu$ &$ 2(\zeta-\nu)$ &$0 $ &$ \sqrt{2}\eta $ &$ 0 $ &$2\kappa $ &$-2\sqrt{2}\mu $ &$ \sqrt{2}\theta $\\
$\phi_{1}\chi_{5}$&$\frac{-2\sqrt{10}}{3}\mu$ &$ \frac{-2\sqrt{2}}{3}(\eta+\mu)$ &$\frac23\kappa $ &$\frac{1}{3}(-8\delta+4\nu) $ &$  0 $ &$
-2\kappa$ &$ \sqrt{2}\eta$ &$  -2\sqrt{2}\xi $ &$  2\sqrt{5}\lambda $ &$  -2(\zeta-\lambda) $ &$ \sqrt{2}\theta $ &$  0$\\
$\phi_{2}\chi_{2}$&${\sqrt{2}}(\theta+3\xi) $ &$ \sqrt{10}\xi $ &$2\sqrt{5}\nu $ &$ 0 $ &$\frac13(\begin{smallmatrix}4\gamma+5\zeta\\-15\nu+3\lambda\end{smallmatrix})$ &$\frac{\sqrt{5}}{3}(5\nu+\lambda)$ &$\frac{-5\sqrt{10}}{3}\xi $ &$\frac{\sqrt{10}}{3}\mu $ &$ -\kappa $ &$ 0 $ &$ 0 $ &$ 0$ \\
$\phi_{2}\chi_{3}$&$-\sqrt{10}\xi $ &$ -\sqrt{2}(\theta+\xi) $ &$ 2(\zeta-\nu) $ &$ -2\kappa$ &$ \frac{\sqrt{5}}{3}(5\nu+\lambda) $ &$
\frac{1}{3}(\begin{smallmatrix}4\gamma+5\zeta+\\+5\nu-\lambda\end{smallmatrix})$ &$\frac{5\sqrt{2}}{3}(\theta+\xi) $ &$ \frac{\sqrt{2}}{3}(\nu-5\eta) $ &$ 0 $ &$ \kappa $ &$ -\sqrt{2}\eta $ &$\sqrt{2}\theta$\\
$\phi_{2}\chi_{4}$&$2\sqrt{5}\nu $ &$2(\zeta-\nu) $ &$ 0 $ &$ \sqrt{2}\eta $ &$\frac{-5\sqrt{10}}{3}\xi $ &$ \frac{5\sqrt{2}}{3}(\theta+\xi)$ &$
\frac{1}{3}(9\alpha-5\delta-2\lambda) $ &$ \frac53\kappa$ &$ 0 $ &$ -\sqrt{2}\eta $ &$ 0 $ &$ -\zeta$\\
$\phi_{2}\chi_{5}$&$0$&$-2\kappa $ &$\sqrt{2}\eta$ &$-2\sqrt{2}\xi $ &$\frac{\sqrt{10}}{3}\mu $ &$ \frac{\sqrt{2}}{3}(\mu-5\eta)$ &$\frac53\kappa$ &$\frac{1}{3}(\begin{smallmatrix}-9\alpha-5\delta\\+10\nu\end{smallmatrix})$ &$0 $ &$\sqrt{2}\theta $ &$ -\zeta $ &$ 0$ \\
$\phi_{3}\chi_{2}$&$\sqrt{2}(\eta+3\mu) $ &$ \sqrt{10}\mu  $ &$0$&$2\sqrt{5}\lambda  $ &$ -\kappa  $ &$ 0 $&$0  $ &$ 0  $ &$\frac{1}{3}(\begin{smallmatrix} -4\beta+5\zeta+\\3\nu-15\lambda \end{smallmatrix})$&$ \frac{-\sqrt{5}}{3}(5\lambda+\nu) $ &$\frac{\sqrt{10}}{3}\xi $&$ \frac{-5\sqrt{10}}{3}\mu $ \\
$\phi_{3}\chi_{3}$&$\sqrt{10}\mu$&$-\sqrt{2}(\eta+\mu)$&$2\kappa$&$ -2(\zeta-\lambda) $&$ 0 $&$ \kappa$&$ -\sqrt{2}\eta $&$\sqrt{2}\theta $&$\frac{-\sqrt{5}}{3}(5\lambda+\nu)$&$\frac{1}{3}(\begin{smallmatrix} -4\beta-5\zeta-\\ \nu+5\lambda\end{smallmatrix})$&$\frac{\sqrt{2}}{3}(5\theta-\xi) $&$\frac{-5\sqrt{2}}{3}(\eta+\mu)$\\
$\phi_{3}\chi_{4}$&$0 $&$ 2\kappa $&$-2\sqrt{2}\mu $&$\sqrt{2}\theta $&$ 0 $&$-\sqrt{2}\eta$&$0 $&$ -\zeta $&$ \frac{\sqrt{10}}{3}\xi  $&$ \frac{\sqrt{2}}{3}(5\theta-\xi) $&$\frac{1}{3}(\begin{smallmatrix}-9\alpha+5\epsilon\\+10\lambda \end{smallmatrix})$&$ \frac53\kappa$\\
$\phi_{3}\chi_{5}$&$2\sqrt{5}\lambda$&$ -2(\zeta-\lambda)$&$\sqrt{2}\eta $&$ 0 $&$ 0 $&$\sqrt{2}\theta
$&$ -\zeta $&$ 0 $&$ \frac{-5\sqrt{10}}{3}\mu $&$\frac{-5\sqrt{2}}{3}(\eta+\mu) $&$\frac53\kappa $&$ \frac{1}{3}(9\alpha+\delta-2\nu)$\\
\midrule[0.5pt]\bottomrule[1.0pt]
\end{tabular}
}
\centering
\caption{The $H_{\rm CMI}$ matrix elements for the case $J^P={\frac12}^{-}$.}\label{13}
\resizebox{\columnwidth}{!}{
\begin{tabular}{|c|ccccccccccccccc|}
\bottomrule[1.0pt]\midrule[0.5pt]
\diagbox{Bases}{Elements}{Bases}&$\phi_{1}\chi_{6}$&$\phi_{1}\chi_{7}$&$\phi_{1}\chi_{8}$&$\phi_{1}\chi_{9}$&$\phi_{1}\chi_{10}$&$\phi_{2}\chi_{6}$&$\phi_{2}\chi_{7}$&$\phi_{2}\chi_{8}$&$\phi_{2}\chi_{9}$&
$\phi_{2}\chi_{10}$&$\phi_{3}\chi_{6}$&$\phi_{3}\chi_{7}$&$\phi_{3}\chi_{8}$&$\phi_{3}\chi_{9}$&$\phi_{3}\chi_{10}$\\
\midrule[0.5pt]
$\phi_{1}\chi_{6}$&$\frac{1}{3}(\begin{smallmatrix}8\alpha-2\zeta\\-4\lambda-4\nu\end{smallmatrix})$&$\frac{4\sqrt{2}}{3}(\nu-\lambda)$&$\frac{2\sqrt{2}}{3}(\theta-2\xi)$
&$\frac{2\sqrt{2}}{3}( 2\mu-\eta)$&$ 0 $&$\sqrt{2}(2\xi-\theta) $&$ -4\xi$&$2(\zeta+2\nu) $&$-2\kappa $&$ 0 $&$ \sqrt{2}(2\mu-\eta) $&$ 4\mu $&$ 2\kappa $&$ -2(\zeta+2\lambda) $&$ 0$\\
$\phi_{1}\chi_{7}$&$\frac{4\sqrt{2}}{3}(\nu-\lambda) $&$ \frac{4}{3}(2\alpha-\zeta) $&$ \frac{4}{3}\xi $&$ \frac{4}{3}\mu $&$\frac{-2\sqrt{3}}{3}\kappa $&$ -4\xi$&$-2\sqrt{2}\theta
$&$-2\sqrt{2}\nu $&$ 0 $&$ -\sqrt{6}\eta$&$ 4\mu$&$ -2\sqrt{2}\eta $&$ 0 $&$ -2\sqrt{2}\lambda$&$ -\sqrt{6}\theta$\\
$\phi_{1}\chi_{8}$&$\frac{2\sqrt{2}}{3}(\theta-2\xi) $&$ \frac{4}{3}\xi $&$ \frac{8}{3}(\epsilon-\lambda) $&$\frac{2}{3}\kappa $&$\frac{-4\sqrt{3}}{3}\mu$&$ 2(\zeta+2\nu) $&$ -2\sqrt{2}\nu $&$0 $&$ \sqrt{2}\eta $&$ 0 $&$ 2\kappa $&$ 0 $&$ 4\sqrt{2}\mu$&$ \sqrt{2}\theta$&$ 2\sqrt{6}\lambda$\\
$\phi_{1}\chi_{9}$&$\frac{2\sqrt{2}}{3}(2\mu-\eta) $&$ \frac{4}{3}\mu$&$ \frac{2}{3}\kappa $&$ -\frac{8}{3}(\nu+\delta) $&$ \frac{-4\sqrt{3}}{3}\xi$&$ -2\kappa $&$0$&$\sqrt{2}\eta$&$ 4\sqrt{2}\xi $&$ 2\sqrt{6}\nu $&$ -2(\zeta+2\lambda) $&$-2\sqrt{2}\lambda $&$ \sqrt{2}\theta$&$ 0 $&$ 0$\\
$\phi_{1}\chi_{10}$&$0 $&$ -\frac{2\sqrt{2}}{3}\kappa $&$-\frac{4\sqrt{3}}{3}\mu $&$ -\frac{4\sqrt{3}}{3}\xi $&$ -8\alpha $&$0 $&$ -\sqrt{6}\eta $&$0 $&$2\sqrt{6}\nu $&$ 0 $&$0 $&$-\sqrt{6}\theta$&$ 2\sqrt{6}\lambda $&$0 $&$ 0$\\
$\phi_{2}\chi_{6}$&$\sqrt{2}(2\xi-\theta)$&$ -4\xi$&$2(\zeta+2\nu) $&$ -2\kappa $&$0 $&$\frac{1}{3}(\begin{smallmatrix}4\gamma-5\zeta-\\10\nu+2\lambda\end{smallmatrix})$&$\frac{2\sqrt{2}}{3}(\lambda+5\nu)$&$\frac{5\sqrt{2}}{3}(\theta-2\xi )$&$\frac{-\sqrt{2}}{3}(5\eta+2\mu )$&$ 0 $&$\kappa $&$ 0 $&$ -\sqrt{2}\eta $&$ \sqrt{2}\theta $&$ 0$\\
$\phi_{2}\chi_{7}$&$-4\xi$&$ -2\sqrt{2}\theta$&$ -2\sqrt{2}\nu$&$0$&$ -\sqrt{6}\eta $&$ \frac{2\sqrt{2}}{3}(\lambda+5\nu) $&$\frac{1}{3}(4\gamma-10\zeta )$&$\frac{10}{3}\xi $&$ -\frac{2}{3}\mu $&$ -\frac{5\sqrt{3}}{3}\kappa $&$0 $&$ 2\kappa $&$ 0 $&$ 0 $&$ \sqrt{3}\zeta$\\
$\phi_{2}\chi_{8}$&$2(\zeta+2\nu) $&$ -2\sqrt{2}\nu $&$ 0 $&$ \sqrt{2}\eta $&$0 $&$\frac{5\sqrt{2}}{3}(\theta-2\xi) $&$ \frac{10}{3}\xi$&$\frac{1}{3}(\begin{smallmatrix}
9\alpha-\\ \epsilon+4\lambda \end{smallmatrix})$&$\frac{5}{3}\kappa $&$ \frac{2\sqrt{3}}{3}\mu $&$ -\sqrt{2}\eta $&$ 0 $&$ 0 $&$-\zeta$&$0$\\
$\phi_{2}\chi_{9}$&$-2\kappa $&$ 0 $&$ \sqrt{2}\eta $&$ 4\sqrt{2}\xi$&$ 2\sqrt{6}\nu $&$\frac{-\sqrt{2}}{3}(5\eta+2\mu) $&$ -\frac{2}{3}\mu $&$\frac{5}{3}\kappa $&$-\frac{1}{3}(9\alpha+5\delta-20\nu)$&$ \frac{-10\sqrt{3}}{3}\xi $&$ \sqrt{2}\theta $&$ 0 $&$ -\zeta $&$ 0 $&$ 0$\\
$\phi_{2}\chi_{10}$&$0 $&$ -\sqrt{6}\eta $&$ 0 $&$2\sqrt{6}\nu $&$ 0 $&$ 0 $&$ -\frac{5\sqrt{3}}{3}\kappa$&$\frac{2\sqrt{3}}{3}\mu $&$\frac{-10\sqrt{3}}{3}\xi $&$ \alpha-3\delta $&$ 0 $&$\sqrt{3}\zeta $&$ 0 $&$0 $&$ 0$\\
$\phi_{3}\chi_{6}$&$\sqrt{2}(2\mu-\eta) $&$  4\mu $&$  2\kappa $&$  -2(\zeta+2\lambda)$&$  0 $&$ \kappa $&$  0
$&$ -\sqrt{2}\eta $&$ \sqrt{2}\theta $&$  0 $&$\frac{1}{3}(\begin{smallmatrix}
-\alpha-3\epsilon-5\zeta\\+2\nu-10\lambda\end{smallmatrix})$&$\frac{-2\sqrt{2}}{3}( 5\lambda+\nu)$&$\frac{\sqrt{2}}{3}(5\theta+2\xi )$&$\frac{5\sqrt{2}}{3}(2\mu-\eta)$&$  0$\\
$\phi_{3}\chi_{7}$&$4\mu $&$ -2\sqrt{2}\eta $&$ 0 $&$ -2\sqrt{2}\lambda$&$-\sqrt{6}\theta$&$ 0 $&$2\kappa
$&$0 $&$ 0 $&$ \sqrt{3}\zeta $&$ \frac{-2\sqrt{2}}{3}(5\lambda+\nu) $&$-\frac{1}{3}(\alpha+3\epsilon-10\zeta $)&$ -\frac{2}{3}\xi $&$ \frac{10}{3}\mu $&$ -\frac{5\sqrt{3}}{3}\kappa$\\
$\phi_{3}\chi_{8}$&$2\kappa $&$ 0 $&$ 4\sqrt{2}\mu$&$\sqrt{2}\theta$&$2\sqrt{6}\lambda$&$ -\sqrt{2}\eta$&$ 0
$&$0 $&$ -\zeta $&$ 0 $&$\frac{\sqrt{2}}{3}(5\theta+2\xi) $&$ -\frac{2}{3}\xi$&$\frac{1}{3}(\begin{smallmatrix}-9\alpha-5\\ \epsilon-20\lambda\end{smallmatrix})$&$ \frac{5}{3}\kappa $&$ -\frac{10\sqrt{3}}{3}\mu$\\
$\phi_{3}\chi_{9}$&$-2(\zeta+2\lambda) $&$ -2\sqrt{2}\lambda $&$\sqrt{2}\theta $&$0 $&$0 $&$\sqrt{2}\theta$&$0
$&$-\zeta $&$ 0 $&$ 0 $&$\frac{5\sqrt{2}}{3}(2\mu-\eta) $&$ \frac{10}{3}\mu $&$\frac{5}{3}\kappa$&$\frac{1}{3}(\begin{smallmatrix}9\alpha+\delta+4\nu\end{smallmatrix})$&$ \frac{2\sqrt{3}}{3}\xi$\\
$\phi_{3}\chi_{10}$&$0 $&$ -\sqrt{6}\theta$&$ 2\sqrt{6}\lambda $&$ 0 $&$ 0 $&$0 $&$
\sqrt{3}\zeta$&$0 $&$0 $&$ 0 $&$0 $&$ -\frac{5\sqrt{3}}{3}\kappa$&$ -\frac{10\sqrt{3}}{3}\mu $&$ \frac{2\sqrt{3}}{3}\xi$&$ \alpha+3\epsilon$\\
\midrule[0.5pt]\bottomrule[1.0pt]
\end{tabular}
}
\end{sidewaystable}
\end{widetext}

\begin{table*}[htbp]
\centering \caption{The $M_{ref}-\langle H_{\rm CMI}\rangle_{ref}$ values of different reference systems in units of MeV. The first lines represent different reference systems for the relevant pentaquark states. The second lines represent the mass values for relevant reference system.}\label{reference}
\renewcommand\arraystretch{1.30}
\begin{tabular}{c|cccccccc}
\bottomrule[1.5pt]
\bottomrule[0.5pt]
&$nncc\bar{c}$&$nncc\bar{b}$&$nnbb\bar{c}$&$nnbb\bar{b}$&$sscc\bar{c}$&$sscc\bar{b}$&$ssbb\bar{c}$&$ssbb\bar{b}$\\ \bottomrule[0.5pt]
Reference &$\Sigma_{c}J/\psi$&$\Sigma_{c}B_{c}^{+}$&$\Sigma_{b}B_{c}^{-}$&$\Sigma_{b}\Upsilon$
&$\Omega_{c}J/\psi$&$\Omega_{c}B_{c}^{+}$&$\Omega_{b}B_{c}^{-}$&$\Omega_{b}\Upsilon$\\ 
System 1&5516.2&8775.5&12104.0&15220.9&5797.7&9056.9&12369.8&15486.7\\ \bottomrule[0.5pt]
Reference &$\Xi_{cc}\bar{D}$&$\Xi_{cc}B$&$\Xi_{bb}\bar{D}$&$\Xi_{bb}B$
&$\Omega_{cc}\bar{D}^{-}_{s}$&$\Omega_{cc}B^{0}_{s}$&$\Omega_{bb}\bar{D}^{-}_{s}$&$\Omega_{bb}B^{0}_{s}$\\ 
System 2&5632.2&8968.2&12079.0&15414.9&5848.3&9176.5&12276.5&15604.7\\
\bottomrule[1.0pt]
&$nnbc\bar{c}$&$nncb\bar{b}$&$ssbc\bar{c}$&$sscb\bar{b}$&$nscc\bar{c}$&$nscc\bar{b}$&$nsbb\bar{c}$&$nsbb\bar{b}$\\ \bottomrule[0.5pt]
Reference &$\Sigma_{c}B_{c}^{-}$&$\Sigma_{c}\Upsilon$&$\Omega_{c}B_{c}^{-}$&$\Omega_{c}\Upsilon$
&$\Xi'_{c}J/\psi$&$\Xi'_{c}B_{c}^{+}$&$\Xi'_{b}B_{c}^{-}$&$\Xi'_{b}\Upsilon$\\ 
System 1&8775.5&11892.4&9057.0&12173.9&5660.4&8919.7&12243.1&15360.1\\ \bottomrule[0.5pt]
Reference &$\Sigma_{b}J/\psi$&$\Sigma_{b}B_{c}^{+}$&$\Omega_{b}J/\psi$&$\Omega_{b}B_{c}^{+}$
&$\Xi_{cc}D_{s}^{-}$&$\Xi_{cc}B_{s}^{0}$&$\Xi_{bb}D_{s}^{-}$&$\Xi_{bb}B_{s}^{0}$\\ 
System 2&8844.8&12104.0&9110.5&12369.8&5730.8&9059.0&12177.5&15505.8\\ \bottomrule[0.5pt]
Reference &$\Xi_{bc}\bar{D}$&$\Xi_{bc}B$&$\Omega_{bc}D_{s}$&$\Omega_{bc}B_{s}$
&$\Omega_{cc}\bar{D}$&$\Omega_{cc}B$&$\Omega_{bb}\bar{D}$&$\Omega_{bb}B$\\ 
System 3&8819.9&12155.9&9022.7&12381.3&5749.7&9085.8&12177.9&15513.9\\ \bottomrule[1.0pt]
\multicolumn{4}{c|}{$nsbc\bar{c}$}&\multicolumn{4}{c|}{$nscb\bar{b}$}\\ \cline{1-8}
Reference&\multicolumn{1}{c|}{$\Xi'_{c}B_{c}^{-}$}&\multicolumn{1}{c|}{Reference}
&\multicolumn{1}{c|}{$\Xi'_{b}J/\psi$}&\multicolumn{1}{c|}{Reference}&\multicolumn{1}{c|}{$\Xi'_{c}\Upsilon$}&\multicolumn{1}{c|}{Reference}
&\multicolumn{1}{c|}{$\Xi'_{b}B_{c}^{+}$}\\
System 1&\multicolumn{1}{c|}{8917.4}&\multicolumn{1}{c|}{System 2}&\multicolumn{1}{c|}{8983.9}&\multicolumn{1}{c|}{System 1}&\multicolumn{1}{c|}{12034.4}&\multicolumn{1}{c|}{System 2}
&\multicolumn{1}{c|}{12243.2}\\ \cline{1-8}
Reference&\multicolumn{1}{c|}{$\Xi_{bc}D_{s}^{-}$}&\multicolumn{1}{c|}{Reference}
&\multicolumn{1}{c|}{$\Omega_{bc}\bar{D}$}&\multicolumn{1}{c|}{Reference}&\multicolumn{1}{c|}{$\Xi_{bc}B_{s}^{0}$}&\multicolumn{1}{c|}{Reference}
&\multicolumn{1}{c|}{$\Omega_{bc}B$}\\
System 3&\multicolumn{1}{c|}{8944.3}&\multicolumn{1}{c|}{System 4}
&\multicolumn{1}{c|}{8954.4}&\multicolumn{1}{c|}{System 3}&\multicolumn{1}{c|}{12272.6}&\multicolumn{1}{c|}{System 4}
&\multicolumn{1}{c|}{12290.5}\\
\bottomrule[0.5pt]
\midrule[1.5pt]
\end{tabular}
\end{table*}

\begin{table*}[htbp]
\centering \caption{The eigenvalues for the $ssbc\bar{Q}$, $nscc(bb)\bar{Q}$ and $nncc(bb)\bar{Q} (Q=c,b)$ systems in units of MeV.
In the $nncc(bb)\bar{Q}$ systems, the $I$ represents the isospin, and
$0$ and $1$ represent the pentaquark states with $I = 0$ and $I = 1$, respectively.}\label{eigenvalue1}
\begin{tabular}{c|cccccc|c|cccc}
\bottomrule[1.5pt]
\bottomrule[0.5pt]
$J^{P}$&\multicolumn{1}{c}{$ssbc\bar{c}$}&\multicolumn{1}{c|}{$sscb\bar{b}$}&\multicolumn{1}{c}{$nscc\bar{c}$}&\multicolumn{1}{c}{$nscc\bar{b}$}
&\multicolumn{1}{|c}{$nsbb\bar{c}$}&\multicolumn{1}{c}{$nsbb\bar{b}$}&\multicolumn{1}{|c|}{$I$}&\multicolumn{1}{c}{$nncc\bar{c}$}
&\multicolumn{1}{c}{$nncc\bar{b}$}&\multicolumn{1}{c}{$nnbb\bar{c}$}&\multicolumn{1}{c}{$nnbb\bar{b}$}\\
\bottomrule[0.7pt]
$\frac{5}{2}^{-}$ &
$\begin{pmatrix}62.5\\51.9\end{pmatrix}$&
$\begin{small}\begin{pmatrix}68.3\\31.8\end{pmatrix}\end{small}$&
$\begin{small}\begin{pmatrix}85.8\\59.0\end{pmatrix}\end{small}$&
$\begin{small}\begin{pmatrix}68.3\\31.8\end{pmatrix}\end{small}$&
$\begin{small}\begin{pmatrix}67.9\\36.9\end{pmatrix}\end{small}$&
$\begin{small}\begin{pmatrix}54.8\\7.4\end{pmatrix}\end{small}$&
\begin{tabular}{c}1 \\ 0\end{tabular}&
\begin{tabular}{c}100.5\\48.5\end{tabular}&
\begin{tabular}{c}82.9\\20.5\end{tabular}&
\begin{tabular}{c}84.0\\29.2\end{tabular}&
\begin{tabular}{c}70.7\\-0.9\end{tabular}
\\
$\frac{3}{2}^{-}$ &
$\begin{small}\begin{pmatrix}84.9\\58.4\\38.0\\33.9\\-6.0\\-31.3\\-91.2\end{pmatrix}\end{small}$&
$\begin{small}\begin{pmatrix}63.5\\50.3\\34.2\\18.3\\-7.4\\-16.7\\-42.7\end{pmatrix}\end{small}$&
$\begin{small}\begin{pmatrix}97.1\\68.7\\49.7\\35.9\\-52.1\\-61.1\\-127.1\end{pmatrix}\end{small}$&
$\begin{small}\begin{pmatrix}90.1\\66.3\\23.7\\14.9\\-3.9\\-37.9\\-94.1\end{pmatrix}\end{small}$&
$\begin{small}\begin{pmatrix}94.0\\55.4\\34.2\\31.9\\-31.9\\-51.0\\-151.6\end{pmatrix}\end{small}$&
$\begin{small}\begin{pmatrix}63.9\\49.5\\41.5\\5.9\\-4.9\\-34.6\\-92.3\end{pmatrix}\end{small}$&
\begin{tabular}{c}1 \\ \\ \\  0\end{tabular}&
\begin{tabular}{c}$\begin{small}\begin{pmatrix}117.3\\81.2\\56.9\\-43.0\end{pmatrix}\end{small}$\\$\begin{small}\begin{pmatrix}36.5\\-68.8\\-161.0\end{pmatrix}\end{small}$\end{tabular}&
\begin{tabular}{c}$\begin{small}\begin{pmatrix}105.1\\84.1\\36.1\\13.6\end{pmatrix}\end{small}$\\$\begin{small}\begin{pmatrix}11.5\\-45.4\\-136.7\end{pmatrix}\end{small}$\end{tabular}&
\begin{tabular}{c}$\begin{small}\begin{pmatrix}115.1\\71.1\\49.2\\-33.0\end{pmatrix}\end{small}$\\$\begin{small}\begin{pmatrix}25.3\\-57.5\\-181.1\end{pmatrix}\end{small}$\end{tabular}&
\begin{tabular}{c}$\begin{small}\begin{pmatrix}83.9\\68.7\\57.7\\12.6\end{pmatrix}\end{small}$\\$\begin{small}\begin{pmatrix}-3.2\\-43.2\\-137.8\end{pmatrix}\end{small}$\end{tabular}
\\
$\frac{1}{2}^{-}$ &
$\begin{small}\begin{pmatrix}125.0\\94.5\\52.1\\11.7\\-8.9\\-39.2\\-90.1\\-138.0\end{pmatrix}\end{small}$
&$\begin{small}\begin{pmatrix}92.9\\63.9\\23.5\\21.5\\-6.6\\-24.6\\-45.5\\-96.2\end{pmatrix}\end{small}$
&$\begin{small}\begin{pmatrix}168.3\\85.0\\27.4\\-14.6\\-48.2\\-83.6\\-86.7\\-240.3\end{pmatrix}\end{small}$
&$\begin{small}\begin{pmatrix}133.0\\55.9\\7.7\\-3.8\\-42.3\\-57.8\\-83.9\\-163.4\end{pmatrix}\end{small}$
&$\begin{small}\begin{pmatrix}118.3\\80.3\\28.6\\12.0\\-34.9\\-50.1\\-92.2\\-194.3\end{pmatrix}\end{small}$
&$\begin{small}\begin{pmatrix}87.7\\62.3\\19.8\\-1.4\\-9.2\\-43.4\\-76.9\\-139.6\end{pmatrix}\end{small}$
&\begin{tabular}{c}1 \\ \\ \\ \\ 0\end{tabular}
&\begin{tabular}{c}$\begin{small}\begin{pmatrix}182.8\\104.4\\3.9\\-62.7\end{pmatrix}\end{small}$\\$\begin{small}\begin{pmatrix}12.8\\-71.0\\-124.3\\-267.2\end{pmatrix}\end{small}$\end{tabular}
&\begin{tabular}{c}$\begin{small}\begin{pmatrix}146.6\\75.8\\27.3\\-21.3\end{pmatrix}\end{small}$\\$\begin{small}\begin{pmatrix}-19.4\\-71.0\\121.9\\-198.8\end{pmatrix}\end{small}$\end{tabular}
&\begin{tabular}{c}$\begin{small}\begin{pmatrix}139.6\\96.9\\28.0\\-32.5\end{pmatrix}\end{small}$\\$\begin{small}\begin{pmatrix}19.6\\-63.6\\-134.1\\-227.7\end{pmatrix}\end{small}$\end{tabular}
&\begin{tabular}{c}$\begin{small}\begin{pmatrix}107.8\\79.8\\37.5\\6.8\end{pmatrix}\end{small}$\\$\begin{small}\begin{pmatrix}-11.1\\-55.7\\-122.1\\-184.6\end{pmatrix}\end{small}$\end{tabular}
\\
\bottomrule[0.5pt]
\midrule[1.5pt]
\end{tabular}
\end{table*}

\begin{table}[htbp]
\centering \caption{The eigenvalues for the $nsbc\bar{Q}$ and  $nncb\bar{Q} (I=1,0) (Q=c,b)$ systems in units of MeV.
In the $nncb\bar{Q}$ systems, the $I$ represents the isospin, and
$0$ and $1$ represent the pentaquark states with $I = 0$ and $I = 1$, respectively.
}\label{eigenvalue2}

\begin{tabular}{c|cc|c|cc}
\bottomrule[1.5pt]
\bottomrule[0.5pt]
$J^{P}$&\multicolumn{1}{c}{$nsbc\bar{c}$}&\multicolumn{1}{c}{$nscb\bar{b}$}&\multicolumn{1}{|c|}{$I$}&\multicolumn{1}{c}{$nnbc\bar{c}$}&\multicolumn{1}{c}{$nncb\bar{b}$}\\
\bottomrule[0.7pt]
$\frac{5}{2}^{-}$ &
$\begin{pmatrix}76.7\\67.7\\46.5\end{pmatrix}$&
$\begin{pmatrix}76.7\\67.7\\46.5\end{pmatrix}$&
\begin{tabular}{c}1 \\ \\ 0\end{tabular}&
\begin{tabular}{c}$\begin{pmatrix}91.3\\83.3\end{pmatrix}$\\37.4\end{tabular}&
\begin{tabular}{c}$\begin{pmatrix}85.9\\71.3\end{pmatrix}$\\8.3\end{tabular}
\\
$\frac{3}{2}^{-}$ &
$\begin{pmatrix}101.1\\77.2\\54.6\\48.5\\40.3\\16.4\\13.6\\-17.1\\-45.3\\-72.3\\-75.7\\-141.4\end{pmatrix}$&
$\begin{pmatrix}78.1\\66.6\\52.3\\35.4\\13.2\\8.9\\1.1\\-11.2\\-23.9\\-41.8\\-85.0\\-93.6\end{pmatrix}$&
\begin{tabular}{c}1 \\ \\ \\ \\ \\ 0\end{tabular}&
\begin{tabular}{c}$\begin{pmatrix}118.8\\95.2\\71.2\\63.2\\33.1\\-2.7\\-54.1\end{pmatrix}$\\$\begin{pmatrix}29.5\\6.0\\-65.4\\-121.9\\-173.1\end{pmatrix}$\end{tabular}&
\begin{tabular}{c}$\begin{pmatrix}94.1\\82.9\\70.4\\52.2\\26.2\\17.9\\-6.2\end{pmatrix}$\\$\begin{pmatrix}2.6\\-21.7\\-49.2\\-131.1\\-138.1\end{pmatrix}$\end{tabular}
\\
$\frac{1}{2}^{-}$ &
$\begin{pmatrix}142.2\\105.7\\73.0\\31.9\\26.2\\5.7\\-18.9\\-36.0\\-54.8\\-70.0\\-88.5\\-119.1\\-138.6\\-192.6\\-238.5\end{pmatrix}$&
$\begin{pmatrix}108.9\\78.6\\44.2\\37.7\\14.0\\-3.4\\-4.2\\-26.6\\-50.7\\-53.9\\-76.4\\-81.8\\-101.1\\-148.0\\-172.7\end{pmatrix}$&
\begin{tabular}{c}1 \\ \\ \\ \\ \\ \\ 0\end{tabular}&
\begin{tabular}{c}$\begin{pmatrix}160.0\\121.5\\93.7\\51.9\\20.6\\1.2\\-50.0\\-100.3\end{pmatrix}$\\$\begin{pmatrix}14.9\\-66.1\\-79.2\\-129.0\\-174.6\\-229.8\\-270.8\end{pmatrix}$\end{tabular}&
\begin{tabular}{c}$\begin{pmatrix}125.6\\93.8\\66.0\\53.4\\34.5\\17.0\\-8.3\\-57.0\end{pmatrix}$\\$\begin{pmatrix}-16.1\\-62.1\\-66.8\\-123.4\\-146.5\\-190.8\\-216.1\end{pmatrix}$\end{tabular}
\\
\bottomrule[0.5pt]
\midrule[1.5pt]
\end{tabular}
\end{table}

\begin{table}[htbp]
\centering \caption{The eigenvalues for the $sscc\bar{Q}$ and $ssbb\bar{Q}$ $(Q=c,b)$ systems in units of MeV. }\label{eigenvalue3}

\begin{tabular}{c|cccc}
\bottomrule[1.5pt]
\bottomrule[0.5pt]
$J^{P}$&\multicolumn{1}{c}{$sscc\bar{c}$}&\multicolumn{1}{c}{$sscc\bar{b}$}&\multicolumn{1}{c}{$ssbb\bar{c}$}&\multicolumn{1}{c}{$ssbb\bar{b}$}\\
\bottomrule[0.7pt]
$\frac{5}{2}^{-}$ &
$\begin{pmatrix}70.9\end{pmatrix}$&
$\begin{pmatrix}53.9\end{pmatrix}$&
$\begin{pmatrix}52.0\end{pmatrix}$&
$\begin{pmatrix}39.2\end{pmatrix}$
\\
$\frac{3}{2}^{-}$ &
$\begin{pmatrix}77.2\\56.2\\14.6\\-78.4\end{pmatrix}$&
$\begin{pmatrix}76.1\\48.4\\-6.4\\-22.9\end{pmatrix}$&
$\begin{pmatrix}73.6\\40.0\\14.9\\-68.9\end{pmatrix}$&
$\begin{pmatrix}45.7\\29.6\\25.8\\-22.3\end{pmatrix}$
\\
$\frac{1}{2}^{-}$ &
$\begin{pmatrix}154.5\\64.9\\-33.8\\-106.4\end{pmatrix}$&
$\begin{pmatrix}120.1\\35.4\\-12.6\\-63.7\end{pmatrix}$&
$\begin{pmatrix}97.6\\63.9\\-3.8\\-67.7\end{pmatrix}$&
$\begin{pmatrix}68.6\\44.8\\2.0\\-25.3\end{pmatrix}$
\\
\bottomrule[0.5pt]
\midrule[1.5pt]
\end{tabular}
\end{table}

\end{document}